\documentclass[10pt,preprint]{aastex}
\usepackage{emulateapj5}

\newcommand{\cmm}{cm$^{-2}$}
\newcommand{\cmmm}{cm$^{-3}$}
\newcommand{\kms}{km~s$^{-1}$}

\newcommand{\lya}{Ly$\alpha$}

\newcommand{\zem}{$z_{em}$}
\newcommand{\zabs}{$z_{abs}$}
\newcommand{\civ}{{\rm C}~{\sc iv}}
\newcommand{\siiii}{{\rm Si}~{\sc iii}}
\newcommand{\siiv}{{\rm Si}~{\sc iv}}

\newcommand{\nv}{{\rm N}~{\sc v}}

\newcommand{\cii}{{\rm C}~{\sc ii}}
\newcommand{\ciii}{{\rm C}~{\sc iii}}
\newcommand{\cv}{{\rm C}~{\sc v}}
\newcommand{\oi}{{\rm O}~{\sc i}}

\newcommand{\alii}{{\rm Al}~{\sc ii}}

\newcommand{\siii}{{\rm Si}~{\sc ii}}
\newcommand{\feii}{{\rm Fe}~{\sc ii}}

\newcommand{\vsh}{$v_{sh}$}
\newcommand{\cf}{$C_{f}$}
\def\lsim{\lower0.3em\hbox{$\,\buildrel <\over\sim\,$}}
\def\gsim{\lower0.3em\hbox{$\,\buildrel >\over\sim\,$}}

\slugcomment{To appear in the Astrophysical Journal}
\shorttitle{Time-Variable Absorption Lines in HS\,1603+3820}
\shortauthors{Misawa et al.}

\begin{document}

\title{Time-Variable Complex Metal Absorption Lines in the Quasar
  HS\,1603+3820\footnotemark[1]}

\footnotetext[1]{Based on data collected at Subaru Telescope, which is
operated by the National Astronomical Observatory of Japan.}

\author{Toru Misawa\altaffilmark{2}, Michael
  Eracleous\altaffilmark{2}, Jane C. Charlton\altaffilmark{2}, and
  Akito Tajitsu\altaffilmark{3}}

\altaffiltext{2}{Department of Astronomy and Astrophysics,
  Pennsylvania State University, 525 Davey Lab, University Park, PA 16802}
\altaffiltext{3}{Subaru Telescope, National Astronomical Observatory
  of Japan, Hilo, HI 96720}

\email{misawa, mce, charlton@astro.psu.edu, tajitsu@naoj.org}

\begin{abstract}

We present a new spectrum of the quasar HS\,1603+3820 taken 1.28 years
(0.36 years in the quasar rest frame) after a previous observation
with Subaru+HDS. The new spectrum enables us to search for time
variability as an identifier of intrinsic narrow absorption lines
(NALs). This quasar shows a rich complex of \civ\ NALs within 60,000
\kms\ of the emission redshift. Based on covering factor analysis,
Misawa et al. found that the \civ\ NAL system at \zabs\ = 2.42--2.45
(System A, at a shift velocity of \vsh = 8,300--10,600 \kms\ relative
to the quasar) was intrinsic to the quasar. With our new spectrum, we
perform time variability analysis as well as covering factor analysis
to separate intrinsic NALs from intervening NALs for 8 \civ\
systems. Only System~A, which was identified as an intrinsic system in
the earlier paper by Misawa et al., shows a strong variation in line
strength ($W_{obs}\approx 10.4~{\rm \AA} \rightarrow 19.1~{\rm
  \AA}$). We speculate that a broad absorption line (BAL) could be
forming in this quasar (i.e., many narrower lines will blend together
to make a BAL profile). We illustrate the plausibility of this
suggestion with the help of a simulation in which we vary the column
densities and covering factors of the NAL complex. Under the
assumption that a change of ionization state causes the variability, a
lower limit can be placed on the electron density ($n_{e} \gsim
3\times 10^{4}~{\rm cm}^{-3}$) and an upper limit on the distance from
the continuum source ($r \leq 6$~kpc). On the other hand, if the
motion of clumpy gas causes the variability (a more likely scenario),
the crossing velocity and the distance from the continuum source are
estimated to be $v_{cross} > 8,000$~\kms\ and $r < 3$~pc.  In this
case, the absorber does not intercept any flux from the broad emission
line region, but only flux from the UV continuum source. If we adopt
the dynamical model of Murray et al., we can obtain a much more strict
constraint on the distance of the gas parcel from the continuum
source, $r < 0.2$ pc.

\end{abstract}


\section{Introduction}

QSO absorption lines are usually classified by the width of their
profiles: broad absorption lines (BALs; FWHM $>$ 2,000 \kms), narrow
absorption lines (NALs; FWHM $<$ 200--300 \kms), and mini-BALs
(intermediate class between BALs and NALs) (Hamann \& Sabra
2003). Because of their large blueshifts and their wide and smooth
line profiles, BALs are plausibly associated with accretion-disk winds
(e.g., Murray \& Chiang 1995; Arav et al. 1995; Proga, Stone \&
Kallman 2000). But BALs appear in only $\sim$ 10~\% of optically
selected quasars (Weymann et al. 1991).

On the other hand, NALs can arise from both intervening absorbers
(e.g., cosmologically intervening galaxies, IGM, and quasar host
galaxies) and intrinsic absorbers (i.e., clouds that are physically
associated with the quasars). A significant fraction of NALs are
thought to be intrinsic to the quasars and may arise in gas that is
outflowing from the central region, based on a statistical analysis
(Richards et al. 1999; Richards 2001). In the spectra of radio-quiet
and flat radio spectrum quasars, intrinsic NALs can appear at a large
shift velocities up to 30,000 \kms\ (Richards et al. 1999). Because
NALs are often not saturated, they give a powerful diagnostic of the
physical conditions of the gas clouds that are physically associated
with the quasars.

In order to understand the quasar central region using NALs, we must
determine which ones are intrinsic to the quasars. We can use several
criteria to separate intrinsic NALs from intervening NALs; e.g., time
variability, high electron density, partial coverage, polarization,
line profile, high ionization state, and high metallicity (Barlow \&
Sargent 1997; Hamann et al. 1997b, and references therein). Among
them, two methods, (i) time variability analysis and (ii) covering
factor analysis, are the most reliable and most frequently applied
(e.g., Barlow et al. 1992; Hamann, Barlow, \& Junkkarinen 1997a;
Barlow \& Sargent 1997; Hamann et al. 1997b; Goodrich \& Miller 1995;
Ganguly et al. 1999,2003; Misawa et al. 2003; Narayanan et al. 2004;
Wise et al. 2004).

In this paper, we apply the above tests to the absorption lines of the
bright quasar HS\,1603+3820 ($z_{em}$=2.542, B=15.9). First discovered
in the Hamburg/CfA Bright Quasar Survey (Hagen et al. 1995; Dobrzycki
et al. 1996), this quasar shows a large number of metal NALs in the
vicinity of the emission redshift. Dobrzycki, Engels, \& Hagen (1999;
hereafter D99) detected 13 \civ\ NALs at 1.965 $<$ \zabs\ $<$ 2.554 in
their intermediate resolution spectrum (R $\sim$ 3000). The number
density of \civ\ Poisson-systems\footnote[4]{Groups of \civ\ NALs that
lie within 1,000 \kms\ each other are classified as a Poisson system,
because these could be produced in the same physical structure (e.g.,
Sargent, Boksenberg, \& Steidel 1988; Steidel 1990; Misawa et
al. 2002).} with rest-frame equivalent width $W_{rest}$ $>$ 0.15 \AA\
is $dN/dz$ $\sim$ 12 at \zabs\ $\sim$ 2.38 in the spectrum of
HS\,1603+3820 (Misawa et al. 2003; M03 hereafter), while the global
average value is $dN/dz = 2.5^{+0.6}_{-0.5}$ at $z_{abs}=2.40$
(Sargent, Boksenberg, \& Steidel 1988; Steidel 1990; Misawa et
al. 2002). Even though the same trends have already been noted
statistically for the class of radio-loud quasars (Foltz et al. 1986),
such an extreme clustering is significant and probably related to the
environment around the quasar (e.g., intrinsic material, a cluster of
galaxies around the quasar, or a proximity effect due to the UV flux
from the quasar).

M03 presented a high-resolution (R$\sim$45,000) spectrum of the quasar
that covered 9 of the 13 \civ\ NALs at \zabs\ $>$ 2.29, although one
of them was apparently not real since it was not detected in that
higher quality spectrum. Using only covering factor analysis, they
concluded that the \civ\ NALs at 2.42 $<$ \zabs\ $<$ 2.45 (System~A in
M03) are intrinsic to the quasar, at velocity shifts\footnote[5]{The
  shift velocity is defined as positive for NALs that are blueshifted
  from the quasar, which is opposite to the definition given in M03.}
from the quasar, \vsh\ = 8,300--10,600 \kms. On the other hand,
Systems B, C, and D at \zabs\ = 2.48--2.55 could not be demonstrated
to be intrinsic, although Systems C and D are close (within 1,000
\kms) to the quasar emission redshift. In fact, one of them (System~D)
is redshifted from the quasar. These absorbers (especially the
redshifted system) could still be intrinsic, but with clouds that
happened to cover the emission source entirely at the time of
observation. Partial coverage is a sufficient, but not a necessary,
condition to demonstrate the intrinsic nature of an absorber.

In this paper, we present a new spectrum of the quasar taken with the
same instrument configuration as our previous spectrum, 1.28 years
after the previous observation (0.36 years in the quasar rest frame)
with the High Dispersion Spectrograph (HDS; Noguchi et al. 2002) on
the Subaru Telescope. Our additional spectrum enables us to search for
time variability of absorption lines as an identifier of intrinsic
NALs, which will not only strengthen our classification results, but
also constrain several parameters of the intrinsic NAL absorbers.

In \S2, we describe the observations and data reduction. The methods
of line identification and line classification are outlined in \S3
and \S4. In \S5, the properties of 8 \civ\ systems, including 4
new systems, are examined in detail, and the possible origins of the
time variability are discussed in \S6. We summarize our results in
\S7. In this paper, we use a cosmology with $H_{0}$=72 \kms
Mpc$^{-1}$, $\Omega_{m}$=0.3, and $\Omega_{\Lambda}$=0.7.

\section{Observation and Data Reduction}

We obtained a spectrum of HS\,1603+3820 with Subaru+HDS on 2003 July 7
(UT). We adopt \zem\ = 2.542 as the systemic redshift of the quasar,
which was estimated by matching the spectrum of D99 to the SDSS
composite spectrum by using the \oi\ $\lambda$1304, \siii\
$\lambda$1307, and \cii\ $\lambda$1335 emission lines (M03). We used a
$0.\!\!^{\prime\prime}8$ slit width (R $\sim$ 45,000) and adopted 2
pixel binning along the slit. The red grating with a central
wavelength of 4900 \AA\ was chosen, which covers the range 3530--4800
\AA\ on the blue CCD and 4980--6180~\AA\ on the red CCD. This
configuration covers all \civ\ NALs identified in D99. We took three
spectra with exposure times of 3,600~s, 3,600~s, and 2,400~s,
respectively. Since the weather conditions were not good at the
beginning of the observation, we used only the last two exposures to
produce the final spectrum. We reduced the data in a standard manner
with the IRAF software\footnote[6]{IRAF is distributed by the National
  Optical Astronomy Observatories, which are operated by the
  Association of Universities for Research in Astronomy, Inc., under
  cooperative agreement with the National Science
  Foundation.}. Wavelength calibration was carried out using the
spectrum of a Th-Ar lamp. Because the blaze profile function of each
echelle order is changing with time, we cannot perform flux
calibration using the spectrum of a standard star. Therefore we
directly fitted the continuum, which also includes substantial
contributions from broad emission lines, with a third-order cubic
spline function. Around heavily absorbed regions, in which direct
continuum fitting is difficult, we use the interpolation technique
introduced in M03. We have already confirmed the validity of this
technique by applying it to a stellar spectrum in M03; the continuum
model is good to 3.3\% or better. After binning every 3 pixels, the
signal-to-noise ratio (hereafter, $S/N$) in our 6,000~s exposure is
$\sim$ 70 per resolution element around $\lambda$ = 5450 \AA. In
Figure~1, we show the normalized spectrum of the region redward of the
\lya\ emission line of the quasar. Echelle order gaps and bad pixels
in the detector cause the defects seen in the spectrum. The spectrum
in the region of the \lya\ forest is not presented because of
low-reliability of continuum fitting, although some strong metal lines
were identified by fitting the continuum locally.

\begin{figure*}
  \includegraphics[scale=0.9]{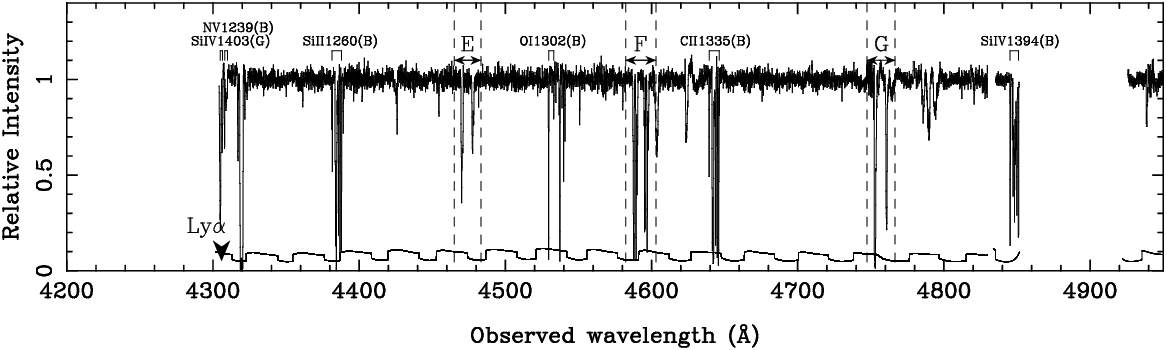}
  \includegraphics[scale=0.9]{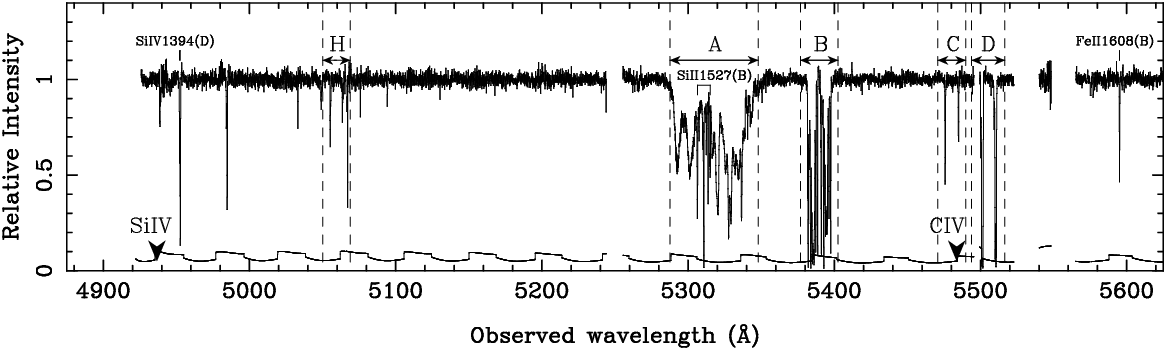}
  \includegraphics[scale=0.9]{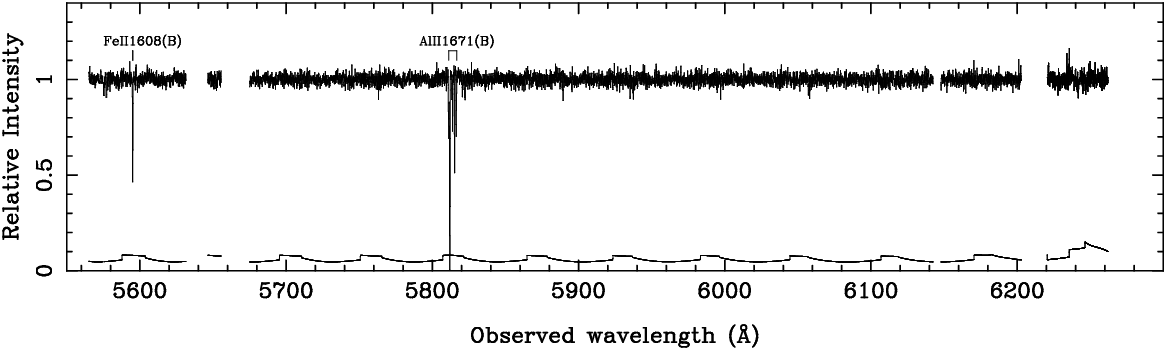}
\caption[Figure 1]{The normalized spectrum of HS\,1603+3820 (binned by
  3 pixels) showing the \civ\ absorption systems as well as other
  metal lines. The quasar emission lines of \lya, \siiv\
  $\lambda$1394, and \civ\ $\lambda$1548 are marked with downward
  arrows. The region blueward of \lya\ emission line is not shown,
  because the spectrum normalization by continuum fitting is not
  reliable in the \lya\ forest. The lower line is the 1$\sigma$ error
  spectrum. The regions affected by either data defects or echelle
  order gaps are masked.}
\end{figure*}

\section{Line Detection and Fitting}

For the present work, we used the line fitting software package {\sc
minfit} (Churchill, 1997; Churchill, Vogt, \& Charlton, 2003), while
we used VPFIT (Webb 1987; Carswell et al. 1987) in M03. The {\sc
minfit} code can fit absorption profiles using not only redshift
($z$), column density ($\log N$ in \cmm), and Doppler parameter ($b$),
but also covering factor (\cf, described below) as a free parameter.

D99 detected 13 \civ\ NALs at 1.88 $<$ \zabs\ $<$ 2.55 in their
intermediate resolution spectrum. In our new spectrum, we identified
11 of the 13 \civ\ NALs in that range. We also found a new \civ\ NAL
at \zabs\ = 2.2653. We, did {\it not} detect two \civ\ NALs found in
D99, at \zabs\ = 2.1762 and 2.5114; the former is unfortunately in the
echelle order gap in our new spectrum, and the latter seems to be a
false detection because it was noted as uncertain in D99 and it is not
detected in either the spectrum of M03 (``the 1st spectrum'',
hereafter) or in the spectrum of this paper (``the 2nd spectrum'',
hereafter). In total, 12 \civ\ NALs were detected in our new spectrum.

Following the method of M03, we divided these \civ\ NALs into 8
systems (Systems A--H in Figure 1). System~A contains 5 \civ\ NALs,
while the other systems have only one NAL, as identified in D99. These
NALs are resolved into multiple components in our high resolution
spectrum (we refer to these as ``doublets'' in the case of \civ, \nv,
and \siiv\ lines and ``lines'' for other single metal lines).  The
first four systems (Systems A--D) had already been identified in
M03. An additional four systems, at \zabs\ = 1.88--2.26 (Systems
E--H), are blueshifted from the other systems and were not covered by
the 1st spectrum. After careful fitting trials with {\sc minfit}, we
found 47 \civ, 14 \siiv, 6 \nv\ doublets, and 54 other single metal
lines associated with the 8 \civ\ systems. We also fitted four systems
in the 1st spectrum again, using {\sc minfit}, and confirmed that the
results with VPFIT in M03 are consistent with the results of {\sc
minfit}.

The fit parameters of these systems are summarized in Table 1 for NALs
already identified in the 1st spectrum and in Table 2 for NALs
detected in only the 2nd spectrum. Column (1) is the metal line
identification (ID) number. Columns (2) and (3) give the observed
wavelength and redshift. Column (4) gives the velocity shift from the
quasar systemic redshift, and column (5), the Doppler
parameter. Columns (6) and (7) give the covering factors and column
densities with 1$\sigma$ Poisson error bars (see discussion in
\S4.1). Columns (2)--(7) in Table 1 are the parameters of the
components detected in the 1st spectrum. The corresponding parameters
for the 2nd spectrum are presented in columns (8)--(13). The same
parameters for newly identified NALs are listed in Table 2. Results of
a fit for the \civ\ doublets in System~A, carried out manually (as
described in \S4.1, below), are given in Table 1. Manual fitting was
necessary because these doublets are so heavily blended with each
other and with other lines that line fitting with {\sc minfit} cannot
be applied. For the same reason, the fit parameters for the \siii\
$\lambda$1526 lines in System~B are evaluated in the same manner, with
fitting errors evaluated with the help of the $\chi^{2}$ test. For all
other doublets, \cf\ is a free parameter.

\section{Identification of Intrinsic NALs}

We describe here two methods of determining that a given NAL is
intrinsic: (a) covering factor analysis and (b) time variability
analysis.

\subsection{Covering Factor Analysis}

The covering factor, \cf, is the line-of-sight coverage fraction of
the absorber over the continuum source and the emission line
region. The effective covering factor is evaluated by considering the
fraction of photons from the background source(s) intercepted by the
absorbing gas. The normalized flux at wavelength $\lambda$ is given
by,
\begin{equation}
R(\lambda) = [1-C_{f} (\lambda)] + C_{f} (\lambda) e^{-\tau (\lambda)}\; ,
\label{eqn:1}
\end{equation}
where $\tau(\lambda)$ is the optical depth at $\lambda$. For a
resonant doublet such as \civ\ $\lambda\lambda$1548,1551, the ratio of
the oscillator strengths of blue and red members is 2, as dictated by
atomic physics. Therefore we can evaluate the covering factors as a
function of velocity for doublets using the following equation:
\begin{equation}
C_{f}(v) = \frac{1+R_{r}^{2}(v)-2R_{r}(v)}{1+R_{b}(v)-2R_{r}(v)}\; ,
\label{eqn:Cf}
\end{equation}
where $R_{b}(v)$ and $R_{r}(v)$ are the residual fluxes of the blue
and red members of the doublets in the normalized spectrum (Hamann et
al. 1997b; Barlow \& Sargent 1997; Crenshaw et al. 1999). In general,
the emission line region and continuum source could have different
covering factors, as discussed by Ganguly et al. (1999).

To strengthen our results, we evaluate the covering factor using the
following methods: (i) the {\it pixel-by-pixel} method -- evaluating
$C_{f}$ for each pixel [using equation~(\ref{eqn:Cf}); e.g., Ganguly
  et al. 1999], and (ii) the {\it automatic fitting} method -- fitting
absorption profiles using $C_{f}$ as well as $\log N$, $b$, and
$z_{abs}$ as free parameters in {\sc minfit} (e.g., Ganguly et
al. 2003). Unfortunately, neither of these two methods is applicable
to System~A because of heavy self-blending (i.e., blending of the {\it
  blue} members of some doublets with the {\it red} members of other
doublets). Therefore we adopt a third method for this system, {\it
  manual} fitting, which we apply only to components 1 and 2, where
blending is not severe (see Figures~2 and 3). In the manual fitting
method we construct a simple model for the absorber with the same free
parameters as in the {\sc minfit} models.  We then vary the free
parameters over a wide range of values in small steps, assessing the
goodness of the fit first by eye, and then by the $\chi^2$ test.

\includegraphics[width=3.3in,bb=30 17 370 500,clip=true]{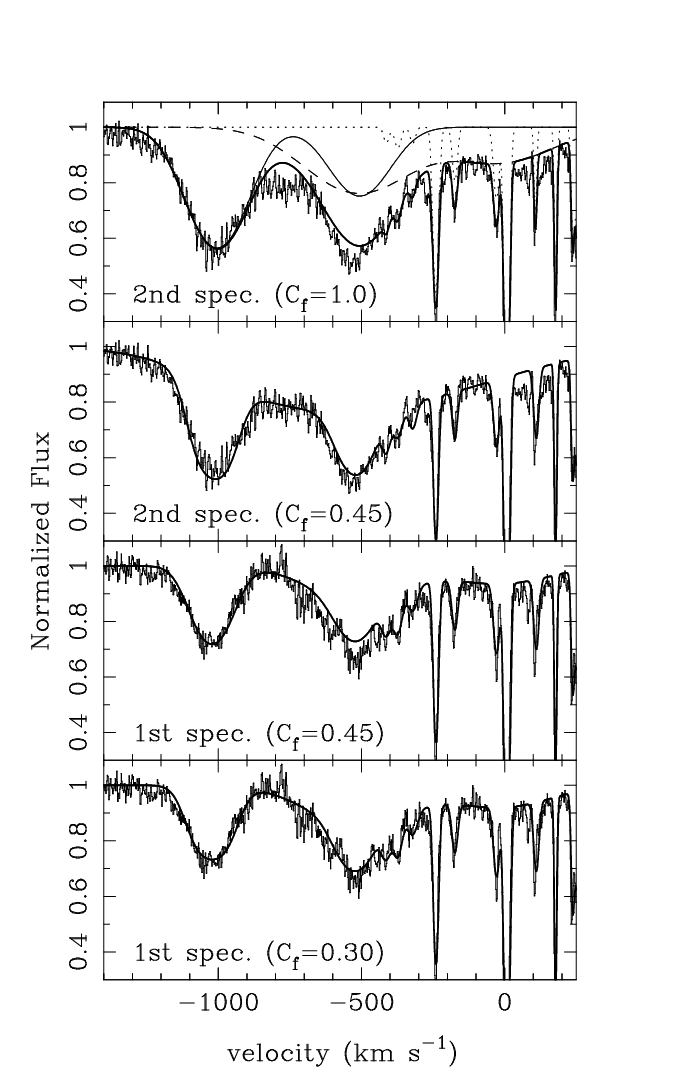}
\figcaption[Figure 2]{{\it First:} The best-fitting model, with fixed
  $C_f=1.0$, of the region around components 1 and 2 of System~A in
  the second-epoch spectrum. Model profiles of each \civ\ doublet and
  \siii\ lines in System~B are also plotted separately (light solid,
  dashed, and dotted lines respectively). The final model is made to
  fit the blue member of the \civ\ doublet. The fit of the red doublet
  member is poor, illustrating that a model with full coverage is not
  acceptable. {\it Second:} The model that provides the best fit to
  the second-epoch spectrum with $C_f$ treated as a free
  parameter. This fit is acceptable, yielding $C_f=0.45$.  {\it
  Third:} The model providing the best fit to the first-epoch
  spectrum with the covering factor fixed to $C_f=0.45$, the value
  required by the second-epoch spectrum (all other parameters were
  left free). The model parameters were adjusted so as to fit the blue
  member of the \civ\ doublet. The fit to the red doublet member is
  poor, however, indicating that the covering factor must have varied
  between the two epochs. Additional details are given in \S4.1 of the
  text. {\it Fourth:} The model that provides the best fit to the
  first-epoch spectrum with $C_f$ treated as a free parameter. This
  fit is acceptable, yielding $C_f=0.3$.}

\includegraphics[width=3.7in,bb=30 17 370 220,clip=true]{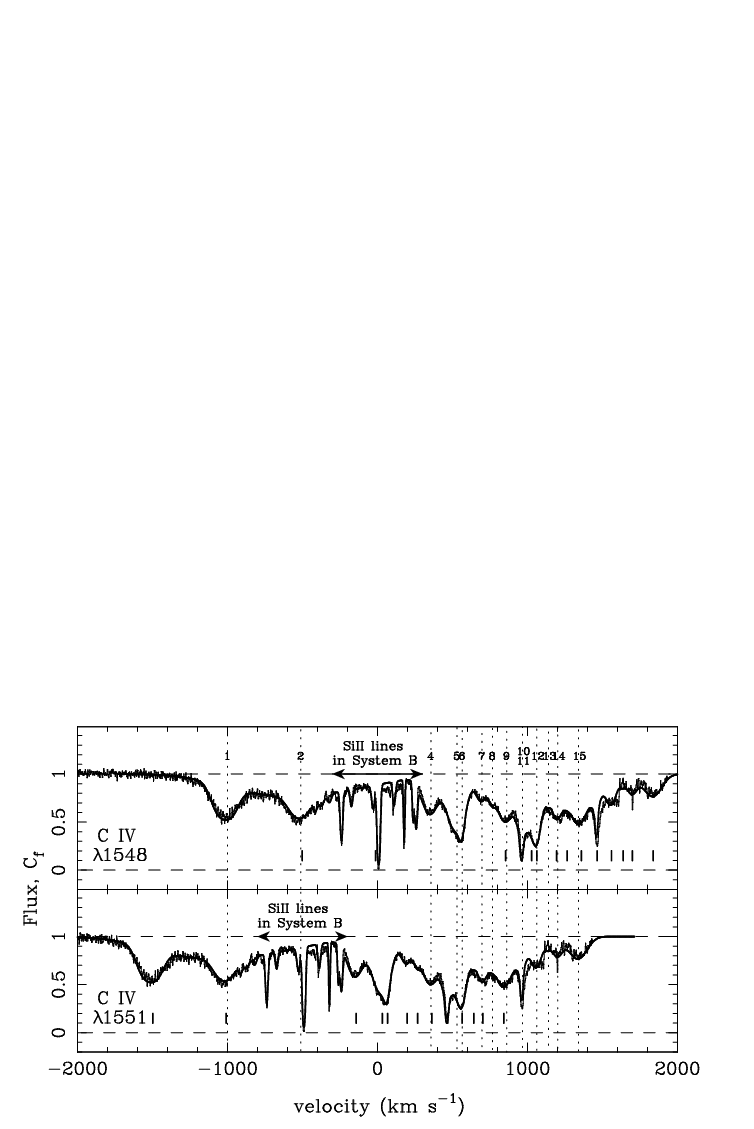}
\figcaption[Figure 3]{Velocity plots of the \civ\ doublets in System~A at
  \zabs\ = 2.4185--2.4455 ($v_{sh}=8,300$--$10,600~{\rm km~s^{-1}}$),
  detected in the 2nd spectrum. Vertical dotted lines with ID numbers
  denote the positions of \civ\ doublets. Smooth lines are models
  fitted manually, using $C_f = 0.3$ for components 1 and 2, and $C_f
  = 1.0$ for the other components.
  The locations of the red and blue members of the 14 \civ\
  doublets are also marked in the top and the bottom panels to illustrate
  the severe blending.}

The $C_{f}$ values with their 1$\sigma$ ``Poisson'' errors, evaluated
by the automatic fitting method, except for the \civ\ doublets in
System~A, are listed in Tables 1 and 2. The Poisson errors are
uncertainties in the model parameters resulting from only Poisson
noise in the spectra.  There are additional sources of uncertainty, of
course, which we discuss in detail below.  In \S5 we present the
spectra and discuss the properties of the 8 \civ\ systems to which we
have applied our analysis. In the case of System~A, we had to fit the
profiles of components 1 and 2 manually. We estimated the Poisson
uncertainties for components 1 and 2 by scanning a range of values
about the best fit for all interesting parameters and looking for a
significant increase in $\chi^2$ (see Lampton, Margon, \& Bowyer
1976). The uncertainties, which are included in Table~1, are
negligible compared to uncertainties resulting from continuum
placement errors or line blending, which we discuss further below.
The ultimate reason for the small Poisson uncertainties in these two
components is that the $S/N$ is high and the lines are relatively
broad, spanning many pixels.

Our derived \cf\ values are subject to a number of additional
uncertainties. One such uncertainty results from line blending. If two
or more components are blended together, a direct estimation of the
\cf\ value by the pixel-by-pixel method [eqn~(\ref{eqn:Cf})] is not
reliable. On the other hand, the results from the automatic fitting
method are still reliable, especially for strong doublets, as long as
they are not affected by self-blending. To evaluate the systematic
uncertainty in the latter method, we performed a line-deblending test
in which we synthesized three artificial \civ\ doublets [$\log (N/{\rm
    cm}^{-2})=13.5,\, 14.0,\, 14.5$; $b=35$~\kms; \cf\ = 0.5]
separated by 100 \kms\ from each other, introduced them into an
artificial spectrum with a $S/N$ similar to what we obtain in
practice, and determined their line parameters with {\sc minfit}. We
repeated this test 500 times. The distribution of resulting covering
factors for the stronger lines is very narrow and centered very close
to the input values; \cf\ = 0.51$\pm$0.06 and 0.51$\pm$0.02 for
doublets with $\log (N/{\rm cm}^{-2})=14.0$ and 14.5. However, the
spread in the measured value of \cf\ is very large for weak doublets
with $\log (N/{\rm cm}^{-2}) =13.5$. The outcome does not change
significantly if we change the input \cf\ values. Thus, the result
from the automatic fitting method is still reliable, at least for
strong doublets with $\log (N/{\rm cm}^{-2}) \geq 14.0$, even if there
is line blending (but not self-blending). However, one must still
attach a systematic error of $\lsim 10$\% to \cf.

Continuum fitting also introduces an uncertainty in the derived values
of \cf\ because it affects the residual flux in the absorbed spectral
regions. We estimated an upper limit to the uncertainty in \cf\
resulting from continuum fitting errors through the following
simulation: we shifted the normalized continuum level around the red
member of the \civ\ doublet by the average noise level of the observed
spectrum [$\sigma(f) \approx 0.015$] while keeping the continuum level
around the blue doublet member fixed and we investigated what the new
value of \cf\ turns out to be, {\it under the assumption that \cf\ was
originally equal to 1}. From this exercise we found that we obtain
\cf\ values 0.98, 0.93, 0.82, 0.56, and 0.13 (instead of 1) for
normalized residual fluxes at the center of blue doublet member of
0.1, 0.3, 0.5, 0.7, and 0.9, respectively. Thus, we conclude that the
\cf\ values for strong doublets (those with a normalized residual flux
of $\le 0.5$) are uncertain by $\lsim 20$\%. We regard this as an
upper limit to the uncertainty because the effect of shifting the
continuum on the value of \cf\ is smaller if (a) we perturb the
continuum around both doublet members, or (b) we assume that the
original value of \cf\ is less than unity. In Table 3, we summarize
upper and lower 1$\sigma$ errors on \cf\ values in the case that \cf\
is less than unity. These were computed by shifting the continuum
level only around the red member. Column (1) gives the \cf\ value and
columns (2) -- (6) are the residual flux of the blue member of the
doublet. We leave the entries blank for unphysical combinations of
\cf\ and $R_b$. These tabulated values are the ones we adopt in our
assessment of uncertainties.

We also considered the uncertainty resulting from the convolution of
the spectrum with the line spread function (LSF) of the
spectrograph. This effect applies only to the pixel-by-pixel method
(since automatic fitting convolves with the LSF) and was originally
explored in the case of narrow doublets (widths comparable to the LSF
width) by Ganguly et al. (1999). We performed the same analysis as
Ganguly et al. but for broader doublets, because most doublets in our
sample are much broader than the LSF. We found that for isolated lines
that are much broader than the LSF, the \cf\ values are evaluated
correctly within a window around the line center whose width is twice
the Gaussian broadening parameter ($2b$). This conclusion depends
slightly on the line strength and width and is not valid if the
doublet members are blended with each other
(``self-blending''). Outside of this window, the \cf\ values are
underestimated, and can reach 0 in the far wings of the line. In these
regions both the numerator and the denominator of
equation~(\ref{eqn:Cf}) are very close to zero, therefore, the \cf\
values obtained with the pixel-by-pixel method are sensitive to the
pixel width, the noise, and the resampling and interpolation schemes
used in the calculation.

Considering all of the sources of error discussed above, the
pixel-by-pixel method yields reliable results for our data in only
very few cases. In the figures presented below, we plot the values of
\cf\ from the pixel-by-pixel method only in the rare cases where their
uncertainties are reasonably small. Otherwise, we rely solely on the
value of \cf\ derived by fitting the line profiles (either manually or
using {\sc minfit}).

\subsection{Time Variability Analysis}

The detection of time variability of absorption line properties such
as strengths, profiles, and shift velocities is another reliable
method for identifying intrinsic NALs. Variability of absorption lines
has been reported for a few Seyfert galaxies (e.g., Kraemer et
al. 2002) and quasars (e.g., Gallagher et al. 2004; Hamann et
al. 1997a).  Recently, larger surveys looking for time variable NALs
have been carried out at $z\sim 2$ (Narayanan et al. 2004) and at $z
\leq 1.5$ (Wise et al. 2004), and they detected variable NALs in 2 of
8 (25\%) and 4 of 15 (27\%) quasars respectively. These fractions,
however, are lower limits, since these studies used low- to
medium-resolution spectra and are thus sensitive only to large
variations.  

To search for spectral variability in HS\,1603+3820, we overplotted
selected regions of the spectra from the two epochs and inspected them
by eye (cf, Hamann et al. 1997a). This simple method is quite adequate
because our spectra have a fairly high resolution and $S/N$ so that
even small changes are obvious to the eye. In summary, we found
substantial variations in System A (Figure 4) but not in any other
absorption system (Figure 5). We describe these changes along with
other results in the following section.  Possible causes of
variability are: (a) a change of ionization state, and (b) a motion of
the absorbing gas across the line of sight. We discuss these
possibilities further in \S6.

\bigskip
\includegraphics[width=3.3in]{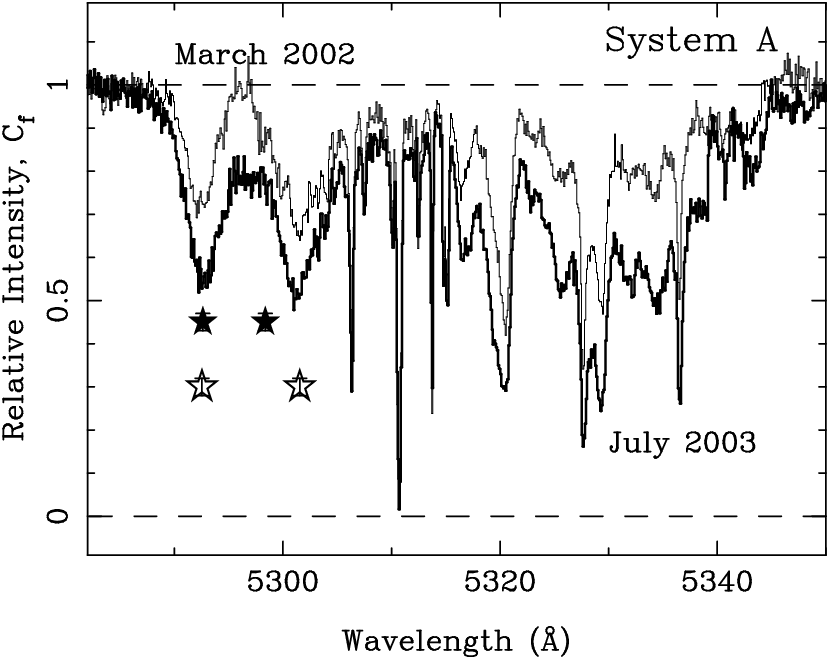}
\figcaption[Figure 4]{Normalized spectrum around \civ\ doublets in
  System~A, in the 1st spectrum (thin histogram) and in the second
  spectrum (thick histogram). There is an obvious change in the
  absorption-line strength between the two epochs. Open and filled
  stars denote the covering factors for the first and second epoch
  respectively (from manual fitting) with 1$\sigma$ errors based 
  on the uncertainty from the line blending only (see discussion of
  additional uncertainties in \S4.1 and \S5 of the text; Poisson
  errors are negligible in comparison).}

\begin{figure*}[b]
\includegraphics[bb=-20 17 400 300]{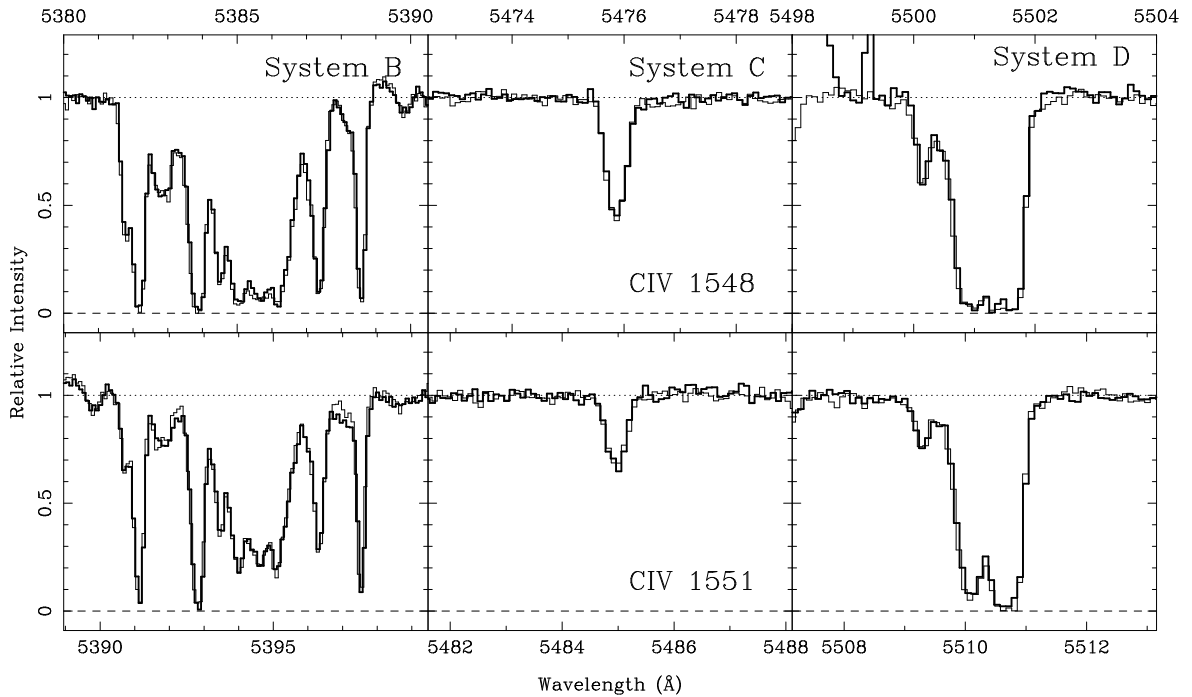}
\caption[Figure 5]{Same as Figure 4, but for System~B, C,
  and D (binned by 10 pixels). The spectra from the two epochs
  are overplotted to illustrate the absence of variability in these
  three systems.}
\end{figure*}

\begin{figure*}[t]
\centerline{
\includegraphics[width=3.in]{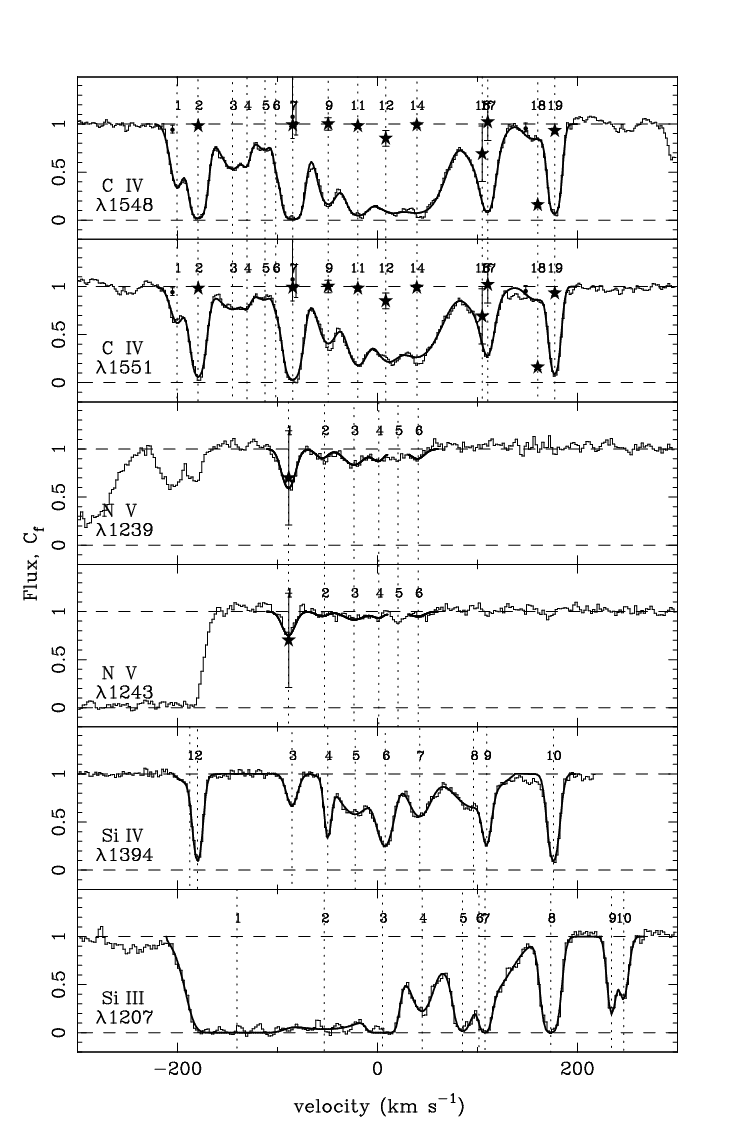}
\includegraphics[width=3.in]{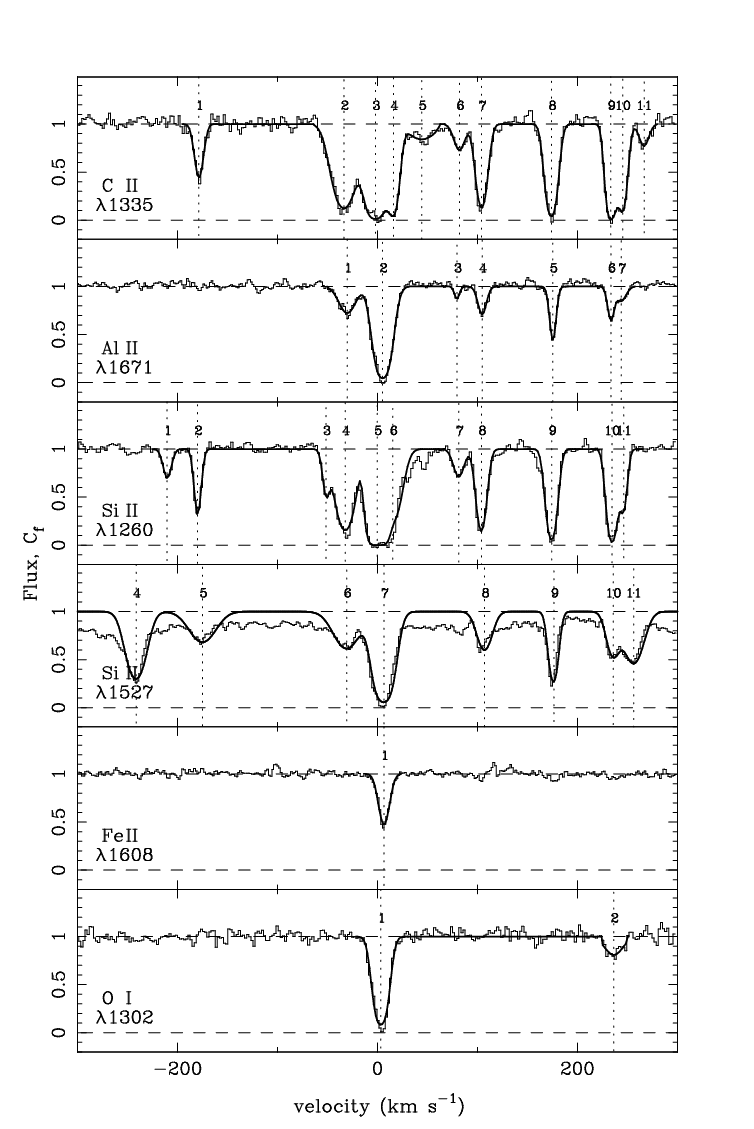}
}
\caption[Figure 6]{Velocity plots of various transitions in System~B
  at \zabs\ = 2.4785 ($v_{sh}=5,400~{\rm km~s^{-1}}$).  Vertical
  dotted lines with ID numbers are positions of absorption
  lines. Smooth lines are model fits by {\sc minfit}. Filled stars
  with error bars are covering factors determined by {\sc
  minfit}. Small dots with error bars denote covering factors
  evaluated by the pixel-by-pixel method. In both cases, the error
  bars represent uncertainties from Poisson noise only; additional
  uncertainties are discussed in \S4.1 of the text.\label{fig:long}}
\end{figure*}

\section{Results: Notes on 8 \civ\ Systems}

We found 47 \civ, 14 \siiv, and 6 \nv\ doublets, as well as another 54
single metal lines, such as \alii, \feii, and \siii, between \zabs\ =
1.88 (\vsh\ = 60,400 \kms) and \zabs\ = 2.55 (\vsh\ = $-$950
\kms). For simplicity, we grouped the 121 components into 8 absorption
systems, which are discussed below. For the first four systems, which
have already been discussed in M03, we performed both covering factor
analysis and time variability analysis. For the other systems, we
applied only covering factor analysis.

\begin{description}

\item[System A ($z_{abs} = 2.4185-2.4455$;] $v_{sh} =$ 8,300--10,600 \kms)
This system has many broad \civ\ doublets and some of them are
self-blended, which makes the line profile very difficult to decompose
into kinematic components. In this case, the results from both the
pixel-by-pixel and the automatic fitting methods are
unreliable. Therefore, we concentrate on the wavelength region at
5290--5315~\AA\ which can be decomposed relatively simply into two
kinematic components of \civ\ plus several narrow \siii\ lines, and
employ the manual fitting method described in \S4.1. If we assume full
coverage, the best-fit model for the red (or blue) doublet member
cannot reproduce the observed spectrum around the blue (or red)
doublet member, as we illustrate in the top panel of Figure~2 using
the 2nd epoch spectrum as an example. Thus, we resort to models with
partial coverage and we find coverage factors of $0.30\pm0.02$ and
$0.45\pm0.02$ for the first and second epoch spectra respectively (all
sources of error included), i.e., we find a $5.3\sigma$ change in
\cf. The uncertainties here are dominated by errors from line blending
as estimated by the line blending analysis in \S\ 4.1; uncertainties
due to continuum fitting (since $C_f = 0.5$ and $R_b\approx 0.5$; see
Table~3) and Poisson noise are negligible in comparison (see \S4.1 and
Table~1).

In the above analysis we assumed the same covering factors for the two
components. We were led to make this assumption for the following
reason: we first fitted the velocity region between $-300$ and
$+250$~\kms\ with a broad \civ\ component only and found that the
required covering factor was the same as that obtained for the narrow
component after fitting a 2-component (broad+narrow) model. After
obtaining the best fit with the composite model we went back and
relaxed the assumption that the two components had the same covering
factors. We explored the consequences by perturbing the covering
factors of the two components about the best fit and adjusting the
column density to compensate. From this exercise we found that, if the
covering factors are not forced to be the same, they can deviate from
the best fit by $\pm0.015$ in the first epoch spectrum and by
$\pm0.03$ in the second epoch spectrum. If we treat these deviations
as an additional error on $C_f$, we find that the change in $C_f$
between the two epochs is $3.4\sigma$, which is still fairly
significant.

To illustrate the change in \cf\ we plot the observed line profiles
from the two epochs in the second and third panels on Figure~2 with
the best fitting models for a fixed value of \cf=0.45 superposed (all
other parameters are free). It is clear from this figure that the
first-epoch profile cannot be fitted adequately using the same value
of \cf\ as the second-epoch profile. We also plot the best fitting
model to the first-epoch spectrum with \cf\ = 0.3 in the bottom panel.

In Figure~3 we present a fit to the entire System A, including the
heavily blended region around 5315--5345~\AA, for which we used the
manual fitting method and assumed a fixed value of \cf=1. The
parameters obtained from this fit are listed in Table~1, although they
should be regarded as tentative because of their large
uncertainties. Deriving the uncertainties for these model parameters
would have been extremely tedious and is not central to the point of
this paper.

We identified 14 \civ\ doublets in this system, while D99 and M03
found 5 and 7 doublets, respectively. \lya, \nv, and \siiv\ lines are
also tentatively detected in our new spectrum. However, we cannot fit
those profiles with {\sc minfit}, because they are very broad
and strong. The continuum fits in these regions are not reliable. We
cannot even apply the interpolation technique described in \S2,
because echelle orders next to a given order are also affected by
\lya\ forest.

For comparison of line profiles between the 1st and 2nd spectra, we
also superimpose the two spectra in Figure 4. The strength of all
doublets, including three narrow doublets, changes dramatically within
1.28 years (0.36 years in the quasar rest frame).  We also found that
both the shift velocity and the line width of the broad component has
changed, while such a change was not noted for component 1. More
specifically, the shift velocity of the broad component changed by
$\sim 210$~\kms\ (the uncertainty in the shift velocity is
$\pm5$~\kms), while the $b$ parameter changed by $\sim 170$~\kms\ (the
uncertainty in $b$ is $\pm8$~\kms).  Unfortunately, we cannot study
these parameters for the other components because of heavy line
blending. The \siii\ $\lambda$ 1527 lines in System~B at 5306--5314
\AA\ do not change between the two observations, which suggests our
interpolation fitting method, described in \S2 and in M03, is
effective. Although M03 suggested that the narrow \civ\ doublets could
be independent of the other lines in System~A based only on the result
of the covering factor analysis, we find here that all 14 doublets
change synchronously, and thus they are probably produced in the same
intrinsic system.  These \civ\ doublets show a sudden increase of
their total equivalent width ($W_{obs}\approx 10.4~{\rm \AA}
\rightarrow 19.1~{\rm \AA}$), which could mean that, in a later stage
of evolution, many narrower components may blend with each other to
make what resembles a BAL profile (Morris et al. 1986; Richards et
al. 2002; M03). Weymann et al. (1991) defined BALs as systems with
continuous absorption, spanning more than 2,000 \kms, within which at
least 10\% of flux must be absorbed. The velocity width of System~A,
$\Delta v$ $\sim$ 2,300 \kms, satisfies this definition, although the
normalized residual flux is more than 0.9 in some regions at the time
of observation.  Some examples of BALs are known to have different
covering factors at different regions in the same system (Arav et
al. 1999; Hutsem\'ekers, Hall, \& Brinkmann 2004). BAL systems also
sometimes have multiple-components, which are analogous to the
components of System~A (Korista et al. 1993; Arav et al. 1999). These
results support the hypothesis that the NAL cluster in System~A could
become a BAL, although additional observations will be necessary to
confirm this. In \S6.3 we use simulations to illustrate the
plausibility of the suggestion that this system could be evolving into
a BAL.

\item[System B $(z_{abs} = 2.4785; ~v_{sh}=5,400~{\rm km~s^{-1}})$]

This strong \civ\ system consists of 15 narrow doublets with $b$
$\leq$ 26 \kms. Some of them can be explained by pure thermal
broadening (i.e., $b \leq 7.4$~\kms\ at $T_{gas}$ = 40,000~K). We
present the velocity plots of 12 metal line transitions, including the
\civ\ and \nv\ doublets, in Figure 6. In cases that {\sc minfit} gives
unphysical \cf\ values (i.e., \cf\ $<$ 0 or $>$ 1), which happens for
some components, probably because of heavy line blending, we assume
\cf\ = 1 to estimate fit parameters. The covering factors of all
doublets except for the component at \zabs\ = 2.4804 (ID 18) are
consistent with full coverage as determined both by profile fitting
and the pixel-by-pixel method. For the latter method, we do not plot
\cf\ values if their error values are larger than 1.0. Component 18 of
\civ\ $\lambda$1551 is likely to be influenced by the unidentified
blend apparent between components 17 and 18, thus the low \cf\
measured for component 18 of \civ\ $\lambda$1551 is
discounted. Component 5 of \nv\ $\lambda$1243 may also be affected by
a blend, since we were unable to obtain a physical solution in this
region for any reasonable continuum fit.

\includegraphics[width=3.3in,bb=30 17 370 220,clip=true]{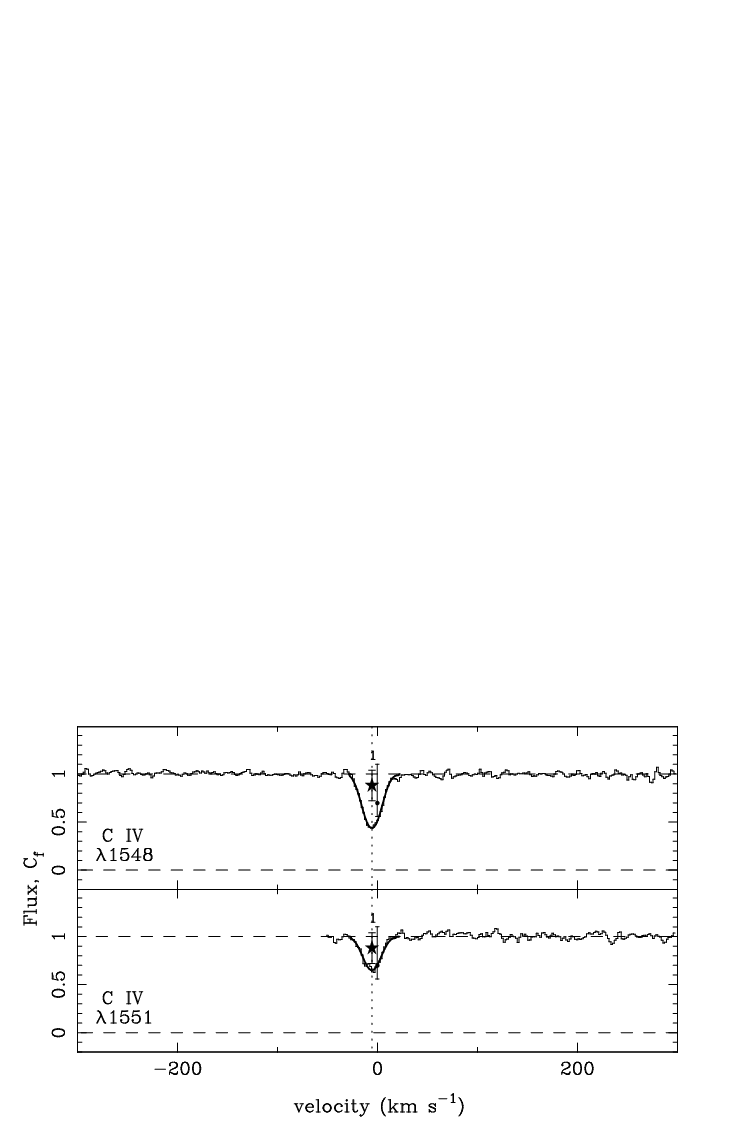}
\figcaption[Figure 7]{Same as Figure 6, but for System~C at \\
  \zabs\ = 2.5370 ($v_{sh}=420~{\rm km~s^{-1}}$).}

\item[System C $(z_{abs} = 2.5370; ~v_{sh}=420~{\rm km~s^{-1}})$]

This is one of two systems within 1,000 \kms\ of the quasar. Besides
the \civ\ doublet, there are no other metal lines detected in the 2nd
spectrum. We find neither partial coverage (Figure 7) nor time
variability (see Figure 5) in spite of the very small velocity shift
from the quasar.

\item[System D $(z_{abs} = 2.5532; ~v_{sh}=-950~{\rm km~s^{-1}})$]

This is another system within 1,000~\kms\ of the quasar, but
redshifted relative to the quasar. This system was resolved into 4
narrow doublets in M03. We are able to analyze only three doublets in
the 2nd spectrum, because the bluest component of \civ\ $\lambda$1548
is affected by a detector defect. Although \siiv, as well as \civ, is
identified in the system, neither shows partial coverage, if we
consider the error in the continuum fit (Figure 8). The \cf\ value of
\civ\ component 1 in the 1st spectrum is small, but the upper
1$\sigma$ error would approach 1.0 once we consider continuum level
error, $\sigma$(\cf) = 0.5 from Table 3.  We also do not find time
variability of the \civ\ doublets (see Figure 5). These results
suggest that the absorber covers the background source entirely in the
direction of this line of sight. Nevertheless, we still cannot rule
out an intrinsic origin for Systems C and D in the vicinity of the
quasar (within 1,000 \kms), especially because a number of intrinsic
NALs have been detected at small velocities from quasars (e.g.,
Ganguly et al. 1999; Narayanan et al. 2004; Wise et al. 2004). We
discuss these systems in \S6.

\includegraphics[width=3.3in,bb=30 17 370 400,clip=true]{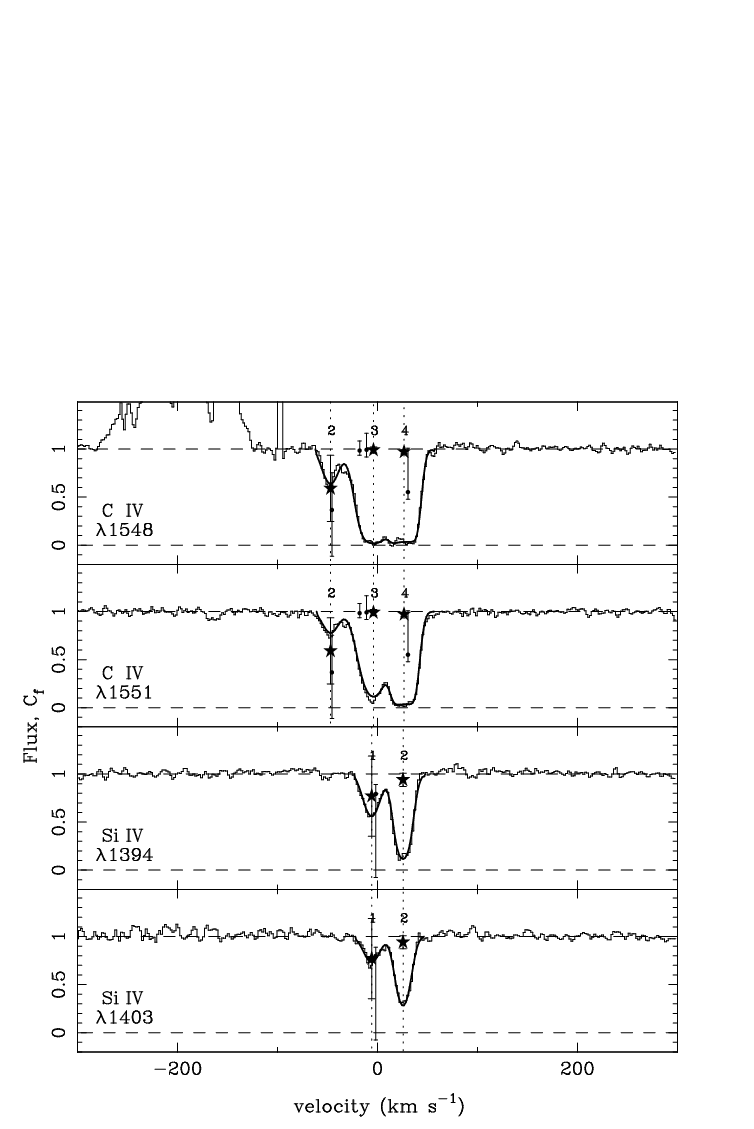}
\figcaption[Figure 8]{Same as Figure 6, but for System~D at \\
  \zabs\ = 2.5532 ($v_{sh}=-950~{\rm km~s^{-1}}$).}

\item[System E $(z_{abs} = 1.8875; ~v_{sh}=60,400~{\rm km~s^{-1}})$]

Two \civ\ doublets with line widths of $b$ = 16 and 27 \kms\ are
identified in the lowest redshift system in this study, which
corresponds to \vsh\ $\sim$ 60,500 \kms\ from the quasar (Figure
9). One of the two doublets (component 2) has a small covering factor,
\cf\ = 0.5$\pm$0.2, and the errors from continuum level
uncertainty ($\sigma$(\cf) = 0.1) and line blending ($\sigma$(\cf)
$\sim$ 0.06) are also very small (in \S\ 4.1). Despite its large
shift velocity, this system could be classified as an intrinsic
system. To confirm this possibility, additional evidence such as time
variability is necessary.

\includegraphics[width=3.3in,bb=30 17 370 220,clip=true]{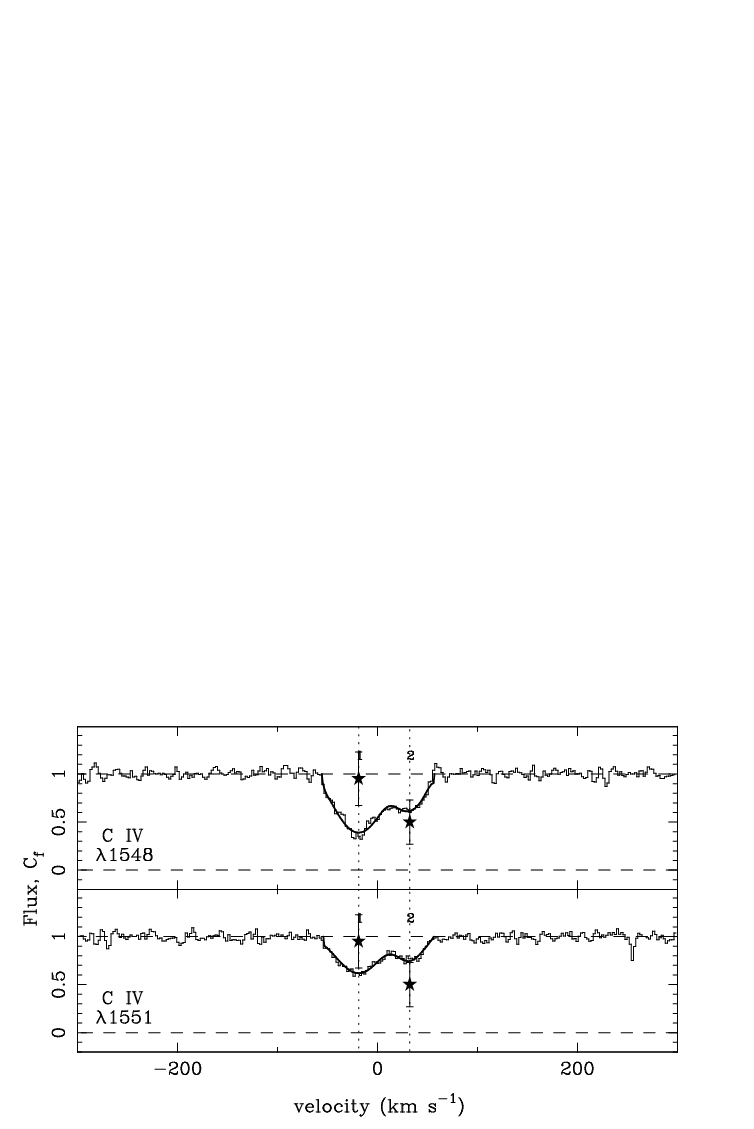}
\figcaption[Figure 10]{Same as Figure 6, but for System~E at \\
  \zabs\ = 1.8875 ($v_{sh}=60,400~{\rm km~s^{-1}}$).}

\item[System F $(z_{abs} = 1.9644; ~v_{sh}=52,800~{\rm km~s^{-1}})$]

Out of eight \civ\ doublets in this system, five show full coverage
(Figure 10). The other three doublets (components 1, 4, and 7) have
small \cf\ values. One of them (component 4) remain a candidate for an
intrinsic doublet, even if we consider both the error sources from 
continuum level uncertainty ($\sigma$(\cf) $<$ 0.1) and line
blending ($\sigma$(\cf) $<$ 0.02) (see discussion in \S4.1).

\includegraphics[width=3.3in,bb=30 17 370 220,clip=true]{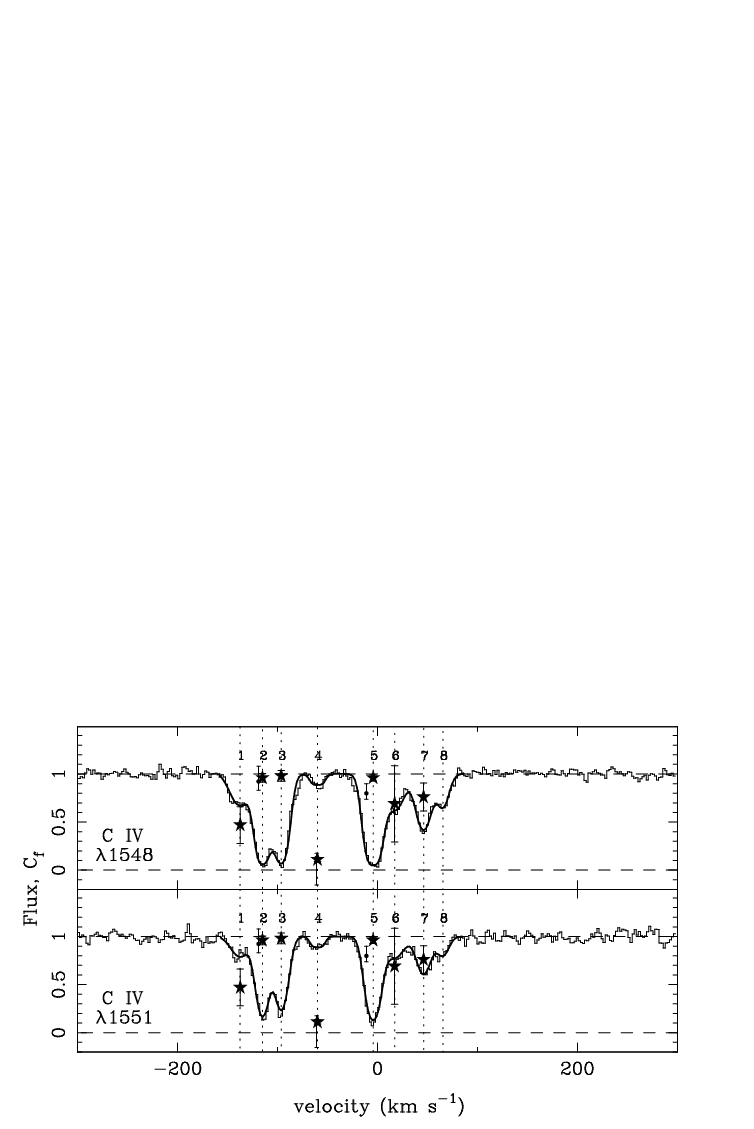}
\figcaption[Figure 10]{Same as Figure 6, but for System~F at \\
  \zabs\ = 1.9644 ($v_{sh}=52,800~{\rm km~s^{-1}}$).}

\item[System G $(z_{abs} = 2.0701; ~v_{sh}=42,600~{\rm km~s^{-1}})$]

This system must also cover the quasar entirely, because the line
centers of the \civ\ $\lambda$1548 transition are black (zero flux)
which cannot happen in the case of partial coverage (Figure 11). The
covering factors of components 1 and 2 are consistent with
unity. Corresponding \siiv\ absorption is apparently detected at
$\Delta v= -30$ and 0~\kms\ (see Figure 11), though the components at
$\Delta v > 50$~\kms\ are \nv\ doublets from System~B. The components
at $\Delta v \leq -50$~\kms\ for \siiv\ are also likely to arise from
another system, because no corresponding \civ\ doublets are
detected. Unfortunately, we cannot evaluate the covering factor of the
\siiv\ doublet, because the \siiv\ $\lambda$1394 line is blended with
the \lya\ forest.

\includegraphics[width=3.3in,bb=30 17 370 300,clip=true]{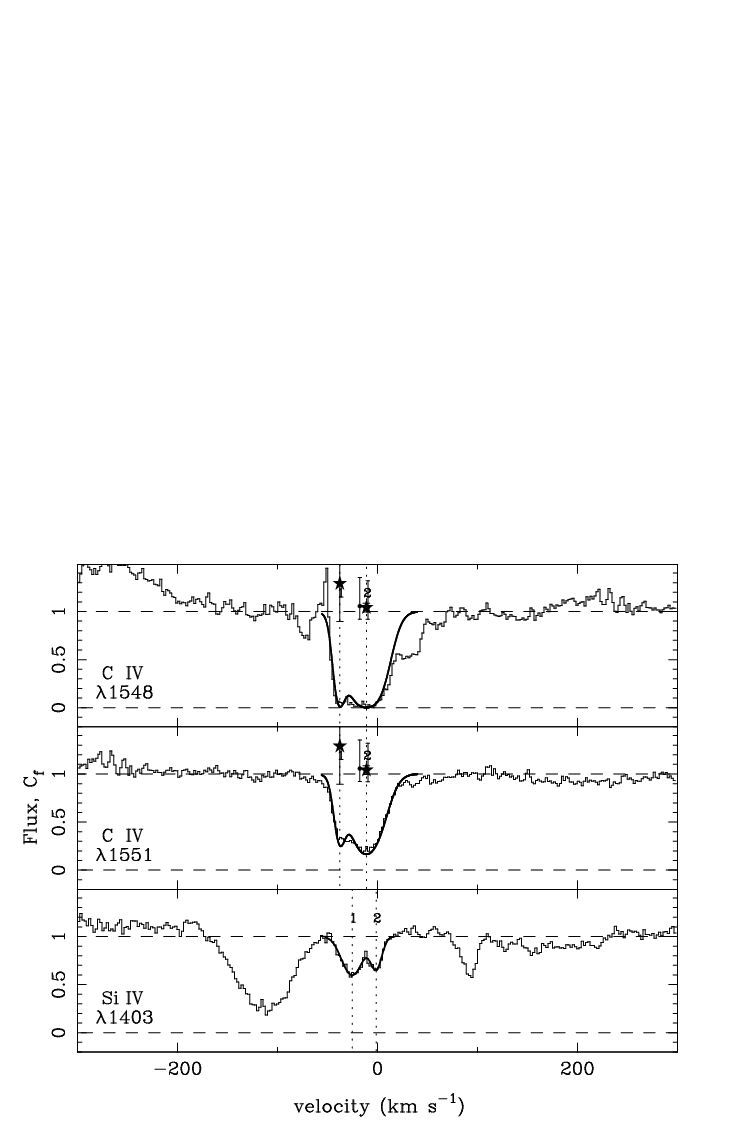}
\figcaption[Figure 11]{Same as Figure 6, but for System~G at \\
  \zabs\ = 2.0701 ($v_{sh}=42,600~{\rm km~s^{-1}}$).}

\item[System H $(z_{abs} = 2.2653; ~v_{sh}=24,300~{\rm km~s^{-1}})$]

A simple \civ\ system with two components, which was not detected in
D99, is identified along with \siiii\ $\lambda$1207 (see
Figure~12). The line center of \siiii\ $\lambda$1207 is well aligned
with \civ\ doublet. Although the covering factor determined by {\sc
  minfit} is smaller than unity ($C_f = 0.5\pm0.3$) for component 1,
we cannot rule out full coverage once we consider the uncertainties
associated with setting the continuum level, $\sigma(C_f) = 0.2$, and
with line blending (see discussion in \S4.1). This component, whose
line width is comparable to the LSF, is affected by the convolution
error. Therefore the \cf\ values from the pixel-by-pixel method are
not reliable. There is a large discrepancy between the model fit and
the data in component 2 of the \civ\ $\lambda$1551 line. This suggests
that the component 2 of the \civ\ $\lambda$1548 line could have been
misidentified as \civ\ or could be affected by continuum fitting
uncertainty.

\includegraphics[width=3.3in,bb=30 17 370 300,clip=true]{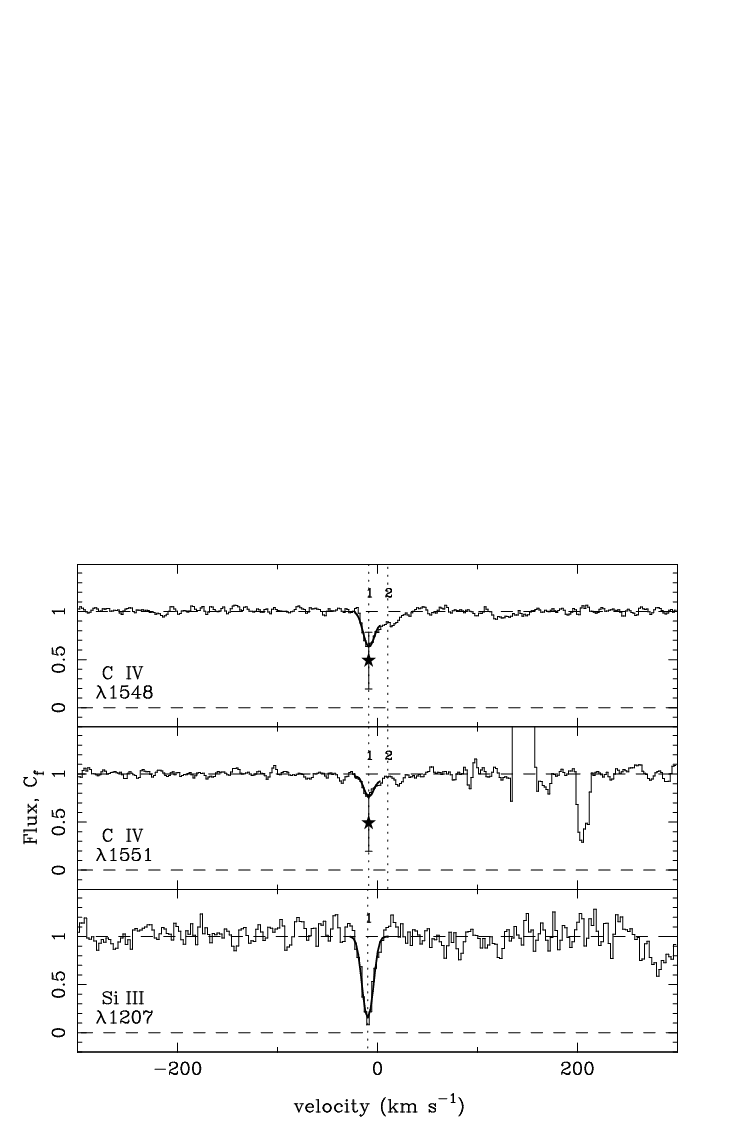}
\figcaption[Figure 12]{Same as Figure 6, but for System~H at \\
  \zabs\ = 2.2653 ($v_{sh}=24,300~{\rm km~s^{-1}}$).}

\end{description} 

\section{Discussion: Origin of Time Variability}

Our main observational result is that System~A shows a change of line
strength within 0.36 years in the quasar rest-frame. No other system
appears to be variable. The fact that System A is variable is
compatible with the fact that it shows the signature of partial
coverage and strengthens the case that the absorber is intrinsic to
the quasar central engine.

There are at least two possible origins of time variability, (a) a
change of the ionization state in the gas clouds, and (b) motion of
the absorbers across the line of sight to the background source(s). We
discuss each of these possibilities below. Neither situation could
arise if the absorbers are intervening clouds, unless they have high
densities or sharp edges (Narayanan et al. 2004). 

\subsection{Change of the Ionization State}

Supposing that the time variability is caused by a change in
ionization state, we can place constraints on the electron density and
the distance from the continuum source, following Hamann et
al. (1997b) and Narayanan et al. (2004). These limits should be
regarded with some caution because of the assumptions involved in
obtaining them. We outline these assumptions below. Changes in the
ionization state of the absorber can be caused by (a) a decrease in
the ionizing flux causing recombination from the observed ionization
state to the next lower state (\civ\ $\rightarrow$ \ciii) or from the
next higher state to the observed state (\cv\ $\rightarrow$ \civ), or
(b) increase in the ionizing flux and subsequent photoionizations
(\ciii\ $\rightarrow$ \civ\ and \civ\ $\rightarrow$ \cv).
Unfortunately, with the available data we cannot assess whether the
ionization state of the absorber of System~A has changed or not.  In
order to do so we would have to resort to photoionization models,
which would need to be constrained by absorption lines from a wide
range of ionic species, which are not included in our spectra. Moreover,
we cannot detect any changes in the observed continuum because the
flux scale of our spectra is not calibrated.
Therefore, we adopt the simplifying assumptions of Hamann et
al. (1997b) and Narayanan et al. (2004), namely: (1) the gas is very
close to ionization equilibrium, (2) changes in the ionizing flux are
small, and (3) \civ\ is dominant ionization state of C.  Under these
assumptions we can use the recombination time scale as a means of
estimating the electron density [eqn.~(1) in Hamann et al. 1997b;
eqn.~(1) in Narayanan et al. 2004; the recombination coefficient
corresponds to a nominal gas temperature of 20,000 K, after Hamann et
al. (1995)].  Using the rest-frame variability time scale ($t_{var}$ =
0.36 years) as an upper limit to the recombination time, we estimate
the minimum electron density to be $n_{e} \gsim 3 \times 10^{4}~{\rm
cm}^{-3}$.  We can also evaluate the maximum distance of the absorber
from the continuum source by adopting the continuum shape of Narayanan
et al. (2004), and using their equation~(3) to connect the ionization
parameter to the bolometric luminosity [the bolometric luminosity of
HS\,1603+3820, based on the prescription of Narayanan et al. (2004) is
$L_{bol}=2.5\times 10^{48}~{\rm erg~s}^{-1}$]. Finally, we must also
assume an ionization parameter of 0.02, which is the optimal value for
\civ\ to be the dominant ionization state of C (Hamann et al. 1995;
1997a). This yields $r \lsim 6$~kpc.
These limits on the electron density and the distance from the central
source are very similar to the results for \civ\ and \nv\ systems seen
in the four quasars ($n_{e} \geq$ 3,400--40,000 \cmmm\ and $r \leq$
1.8--7~kpc) studied by Narayanan et al. (2004).

In conclusion, we note that regardless of changes in ionization state,
however, we can conclude with reasonable confidence that the absorber
is moving across the line of sight because the covering factor of the
background source(s) is changing. In the next section we consider the
implications of the variable covering factor in detail.

\subsection{Motion of the Absorber}

In the case of motion of the absorbers, the variability time scale
could correspond to (i) the crossing time of clumpy gas across the
background source or (ii) the time scale on which the internal motion
of the gas cloud changes the column density along the line of sight
(e.g., Hamann et al. 1997a; Wise et al. 2004). 

The variability time scale of System~A is very short ($\sim$ 130
days), suggesting that the observed changes could not be produced by
internal motions of the gas cloud, as we will illustrate later in this
section.  Some intrinsic NALs detected in Narayanan et al. (2004) and
Hamann et al. (1997a) also show very short variability time scales
($t_{var}$ = 0.28--0.40 years) in the quasar rest frame. In these
systems, the clumpy clouds could be moving across the line of sight to
the continuum source. If this is the case for System~A, we can derive
some constraints on the size, crossing velocity, and distance of the
absorber from the UV-emitting source using a simplified model as
follows.

At first, we assume that both the continuum source and BLR are
background sources (e.g., Ganguly et al. 1999). In this case, we can
estimate the size of the BLR to be $R_{BLR}\sim 3$~pc by using the
empirical relation between BLR size and the quasar luminosity that was
found through reverberation mapping analysis [eqn~(A3) in Vestergaard
2002]. To apply this relation, we used the monochromatic continuum
luminosity derived from the $B$ magnitude of HS\,1603+3820, measured by
D99. This luminosity is strictly an upper limit because at the
redshift of the quasar the $B$ filter includes a contribution from the
Ly$\alpha$ and \ion{N}{5} emission lines; nevertheless, the luminosity
is overestimated only by a factor of a few.

Supposing that the absorbers have sharp edges, we estimate the
crossing velocities as follows:
\begin{equation}
v_{cross} > \left[C_{f}(2) - C_{f}(1)\right]\times\frac{R_{BLR}}{t_{var}}\; ,
\label{eqn:Vcross}
\end{equation}
where $v_{cross}$ is the crossing velocity perpendicular to our line
of sight (in the context of a disk-wind model this is parallel to the
surface of the accretion disk), \cf(1) and \cf(2) are covering factors
in the 1st and 2nd spectra, and $t_{var}$ is the time interval between
the two observations. Since $t_{var}$ is really an upper limit to the
variability time scale, equation~(\ref{eqn:Vcross}) gives a lower limit to
the crossing velocity.
The crossing velocities of the two \civ\ components in System~A
(components 1 and 3 in the 1st spectrum, or components 1 and 2 in the
2nd spectrum), for which we can evaluate reliable \cf\ values in both
spectra, would be $v_{cross} > 10^6$~\kms. This model is implausible,
because the estimated speeds are much larger than the speed of
light. Thus, we must conclude that the absorber covers primarily the
UV continuum source and only a small part of the BLR. In fact, since
the intensity of the broad \civ\ emission line at the location of
System A is only about 10\% of the continuum, the absorber giving rise
to System~A need not cover the BLR at all.

Therefore, we estimate the velocities by assuming that only the
continuum source lies behind the absorbing clouds (e.g., Wise et
al. 2004). We take $5\; R_{S} = 10\; GM_{BH}/c^2$ (where $R_{S}$ and
$M_{BH}$ are the Schwarzschild radius and the mass of black hole) as
the continuum source size within which most of the UV continuum
radiation is expected to originate. The central black hole mass is
estimated by using an empirical relation that connects it to the UV
luminosity and the width of the \ion{C}{4} emission line [eqn~(8) in
Vestergaard (2002)]. The UV luminosity was estimated as described
above, and the FWHM of the \ion{C}{4} emission line (7,500~\kms) was
measured from the spectrum of D99. The resulting black hole mass is
$4\times 10^{10}~{\rm M}_{\odot}$, but we note that this value should
be regarded with some caution because (a) the scatter about the
empirical relation we have used can be as high as an order of
magnitude, and (b) the luminosity we have adopted is strictly an upper
limit to the true value. Therefore, we believe that the black hole
mass can be as low as a few $\times 10^9~{\rm M}_{\odot}$. This model
produces reasonable values of the crossing velocities, namely
$v_{cross} > 8,000$~\kms, using the higher black hole mass.  We may
also use the black hole mass to estimate an upper limit to the
distance of the absorber from the background source, by requiring that
the observed shift velocity does not exceed the escape velocity at
that radius. From this consideration we obtain $r\lsim 3$~pc.

The above estimate of the crossing velocity can be used to argue
against internal motions of the absorber as the origin of the change
in the covering factor. Assuming that the speed of internal motions is
comparable to the speed of sound, $c_s\approx 10\; T_4^{1/2}$~\kms\
(where $T_4$ is the temperature in units of $10^4$~K), the gas must be
very hot to allow such fast internal motions. In particular, by
setting $v_{cross} = c_s$ we obtain $T_{gas}\sim 6\times 10^8$~K,
which is incompatible with the observed ionic species. We note, in
conclusion, that using the size of the UV continuum source from the
previous paragraph and the speed of sound for a $10^4$~K gas, a
typical variability time scale due to internal gas motions would be
of order 100 years.

We can derive more stringent, albeit more speculative, constraints on
the location of the absorber, if we adopt the dynamical model of
Murray et al. (1995) for the radial and azimuthal motion of parcels of
gas in an accretion-disk wind. Following this model, we assume that
the velocity of the gas in the radial direction follows $v_{r}(r)
\simeq v_{\infty}(r_{f})(1-r_{f}/r)$, where $r_{f}$ is the launch
radius of a specific gas parcel and $v_{\infty}(r_{f})$ is its
terminal speed. Combining this with conservation of angular momentum
and the expression for the terminal speed given in Chiang \& Murray
(1996) we obtain the following relation for the radius of the gas
parcel relative to its launch radius:
\begin{equation}
{r\over r_{f}} \approx 
0.2\; \left[{v_{r}(r)\over v_{\varphi}(r)}\right] + 1\; ,
\label{eqn:4}
\end{equation}
where $v_{\varphi}(r)$ is the azimuthal speed of the parcel. To
connect this expression to the observations, we identify the azimuthal
velocity with the crossing velocity of the continuum source inferred
from observations, i.e. $v_{\varphi}(r)=v_{cross}$. Moreover, we can
take the velocity of System~A to be the projection of the
radial velocity along the line of sight, i.e., $v_{r}(r) \sin
i=v_{Sys~A}$. If the inclination of the accretion disk is not very
small ($\sin i > 0.2$), we can use equation~(\ref{eqn:4}) to obtain an
upper limit on $r/r_{f}$. This limit will depend directly on the
adopted mass of the black hole, since that determines the inferred
crossing velocity. In view of the caveats associated with the
estimated black hole mass, we choose to use the lowest plausible
value, namely $~10^9~{\rm M}_{\odot}$, which yields the most generous
upper limit on the radius of the absorber, namely $r/r_f \lsim 10$.
The launch radius of the wind is expected to be within the inner,
radiation-pressure supported part of the accretion disk, which for our
inferred black hole mass and assuming an accretion rate close to the
Eddington limit, is $5\times 10^{16}$~cm (Shakura \& Sunyaev 1973;
this is rather insensitive to the black hole mass). Thus the
constraint on the radius translates to $r \lsim 0.2$~pc. We also note
that the conclusion that $r/r_{f} < 10$ is in good agreement with the
simulations of Proga et al. (2000), which show the wind fragmenting
into filaments fairly close to its launch radius.\footnote[7]{If we had
chosen the higher black hole mass of $4\times 10^{10}~{\rm
M}_{\odot}$, we would have obtained $r/r_{f} \sim 1$ and $r \lsim
0.02$~pc.}

\subsection{Evolution to a BAL}
The total equivalent width of System~A has doubled within 0.36 years
in the quasar rest frame. This surprising growth is probably caused by
(i) increase of column densities, and/or (ii) increase of the covering
factors of the \civ\ doublets in the system, although the Doppler
parameters could also be partially responsible. If the equivalent
width continues to increase steadily, System~A would satisfy all the
requirements for a BAL classification.

In order to predict future profiles, we synthesize spectra by changing
the column densities or the covering factors of the \civ\ doublets in
the system. Unfortunately, we cannot evaluate the column density and
covering factor of each doublet exactly in the observed spectrum,
since the \civ\ doublets are heavily blended with each other or with
\siii\ lines from System~B. Therefore, we convert the column densities
of the \civ\ doublets in the 2nd spectrum in Table 1 (evaluated
assuming \cf\ = 0.45 or 1.0) to values assuming \cf\ = 0.7
(corresponding to the depth of the deepest \civ\ component; we adopt
this value for all components for simplicity), and use them below.
We present simulated line profiles which are synthesized by
multiplying the column densities by 0.5, 1.0, 2.0, and 5.0 (Figure 13;
here, covering factors are fixed), or by changing the covering
factors, \cf\ = 0.3, 0.5, 0.7, and 0.9 (Figure 14; here, column
densities are fixed). Thick lines are the observed profiles, while
thin lines are synthesized models. For three models (i.e.,
$N_{2}/N_{1}$ = 2.0, 5.0, or \cf\ = 0.9), the continuum flux is
absorbed by at least 10\% over a velocity range of $>$ 2,000 \kms,
which satisfies the definition of a BAL. If the Doppler parameters
increase simultaneously, the blended \civ\ components make a smoother
profile, whose appearance would be similar to those of BALs.  With
additional spectra taken at regular intervals, we could probe the
origin of time variability further (i.e., distinguish between a change
of column density and a change in covering factor) by comparing the
observed line profiles with our simulations.

\includegraphics[width=3.3in]{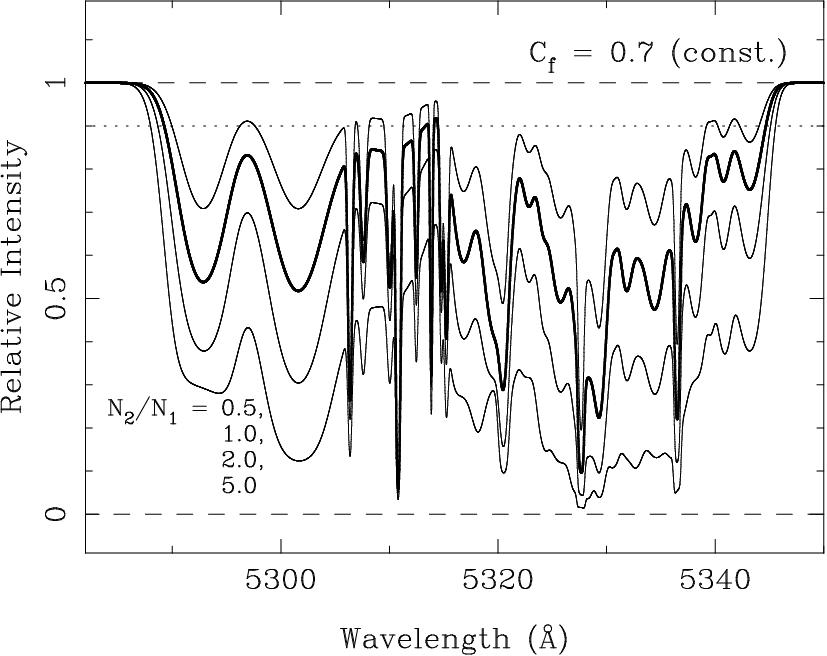}
\figcaption[Figure 13]{Model profiles of System~A, synthesized by
  multiplying column densities of the \civ\ components by factors of
  $N_{2}/N_{1}$ = 0.5, 1.0, 2.0, and 5.0 (top to bottom). The covering
  factors are assumed to be constant (\cf\ = 0.7 in all cases). The
  original column densities were estimated, assuming that \cf\ = 0.7
  for all \civ\ doublets in the observed spectrum (except for the
  \siii\ lines in System~B). The thick line corresponds to the
  original (observed) profile, while thin lines are model profiles for
  different column densities. The total velocity width of the system
  is about 2,300 \kms. Therefore, the model profile is classified as a
  BAL, if its normalized flux is weaker than 0.9 (horizontal dotted
  line) throughout the system. Two of the models ($N_{2}/N_{1}$ = 2.0
  and 5.0) satisfy all the BAL criteria.}
\bigskip
\bigskip
\includegraphics[width=3.3in]{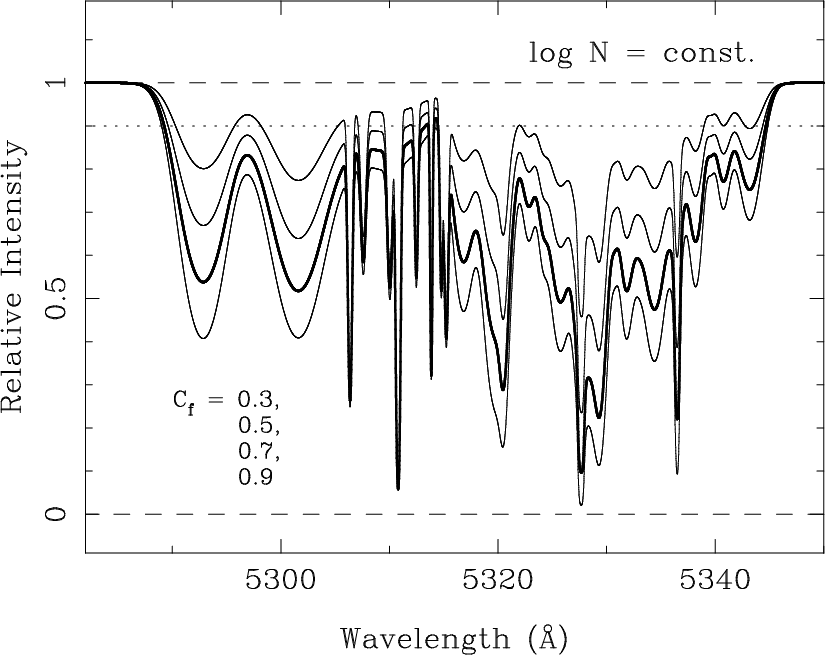}
\figcaption[Figure 14]{Same as Figure 13, but synthesized by changing
  covering factors, \cf\ = 0.3, 0.5, 0.7, and 0.9 (top to bottom).
  Column densities are fixed. One of these models (\cf\ = 0.9)
  satisfies the criteria for a BAL.}

\subsection{Physical States of System C and D}

In Systems C and D, we detect neither time variability nor partial
coverage with our spectra taken over a 1.28-year interval of
observation, even though these would appear to be promising candidates
for intrinsic systems because of their small shift velocities.

There are at least three candidates for the origin of these
systems. They can be produced in either (i) the quasar host galaxy,
(ii) intervening galaxies in the vicinity of the quasar that are
members of the same cluster (or group) of galaxies, or (iii) gas
clouds intrinsically associated with the quasar that {\it appear} to
be ejected (System~C) or infalling (System~D); these gas parcels can
be part of a high altitude stream in an accretion disk wind (which may
be circulating) as simulated by Proga et al. (2000).

However, since we do not see the hallmarks of intrinsic absorption in
either of these two systems (e.g., partial coverage or time
variability), we cannot prove that they are intrinsic, nor can we rule
this out. We find it unlikely that System~D is made up of undisturbed
gas in the host galaxy because of its large velocity shift. The fact
that the absorption troughs in System~D are black, suggests that this
absorber covers both the continuum source and the broad-emission line
region almost completely, i.e., its transverse size could be a few
pc (cf, \S6.2). This large size could also be responsible for the lack
of variability, if this system is intrinsic.

\subsection{Other Potentially Intrinsic Systems}

Among the 8 \civ\ systems, four (Systems E -- H) have been studied
based on only covering factor analysis. In cases where the measured
\cf\ value is small, the errors of \cf\ from continuum level
uncertainty, as listed in Table 3, are small, and
such components can be classified as intrinsic doublets: component
2 in System E (\cf\ = 0.5$\pm$0.2) and component 4 in System F (\cf\ =
0.1 with $\sigma$(\cf) smaller than 0.1) (here errors from the line
blending are negligible). These systems are potentially interesting
because of their high velocities relative to the quasar: \vsh\ $\sim$
60,400 \kms\ (System E) and 52,800 \kms\ (System F). Therefore, it is
very important to monitor them further for time variability. 
We already know of several examples of high-velocity
systems, which are variable. These include the systems 
in Q0151$+$048 with \vsh\ $\sim$ 28,000 \kms, 
in PG0935$+$417 with \vsh\ $\sim$ 52,000 \kms\ (Narayanan et al. 2004), 
in Q2343$+$125 with \vsh\ $\sim$ 24,000 \kms\ (Hamann et al. 1997a), and 
in PG2302$+$029 with \vsh\ $\sim$ 56,000 \kms\ (Jannuzi 2002).

\section{Conclusions and Future Work}

We observed the quasar HS\,1603+3820 for a second time with
Subaru+HDS after an interval of 1.28 years from the previous
observation, in order to perform time variability analysis for the
\civ\ systems at \vsh\ = $-$950 \kms\ -- 10,600 \kms\ from 
the quasar. We also applied
covering factor analysis using two methods, the pixel-by-pixel method
and Voigt profile fitting. Our main results are:

\begin{itemize}
\item[(1)]{Only System~A (\vsh\ $\sim$ 8,300--10,600 \kms), which was
  identified as an intrinsic system by covering factor analysis in
  M03, shows strong time variability of the line strength and covering
  factor.}
\item[(2)]{If a change of the ionization state causes the variability
  in System~A, we can place constraints on the electron density
  ($n_{e} \gsim 3 \times 10^{4}~{\rm cm}^{-3}$) and the absorber's
  distance from the continuum source ($r < 6$~kpc).}
\item[(3)]{If gas motion across the background UV source causes the
  variability in System~A, the crossing velocity and the distance from
  the continuum source are estimated to be $v_{cross} > 8,000$~\kms\
  and $r < 3$~pc, respectively. In this case, the BLR cannot
  contribute significantly to the background UV flux (in fact, it need
  not contribute at all). If we adopt the dynamical model of Murray et
  al. (1995), we obtain an additional relation for the radius of the
  gas parcel relative to its launch radius, which gives a more strict,
  although model-dependent, constraint of $r < 0.2$ pc.}

\item[(4)]{We argue that the appearance of System~A would
  be fairly similar to that of a BAL system, if either the column
  densities of the \civ\ lines increase by factors of $\geq$ 2, or if
  their covering factors become larger than 0.9.}

\item[(5)]{Systems C and D (\vsh\ $\sim$ 420 \kms\ and $-$950 \kms) do
  not show any time variability, in spite of their small velocity
  distance from the quasar. They could still be intrinsic systems 
  but a more stringent test will have to await future observations}

\item[(6)]{Two systems (System~E and F; \vsh\ = 60,400 and 52,800 \kms)
  show suggestive evidence of an intrinsic origin 
  based on only covering factor
  analysis. To confirm this, we would need other evidence for
  intrinsic NALs, such as time variability.}

\end{itemize}

In order to investigate the nature of System~C further, there are at
least two possible future observations. If we observe this quasar
repeatedly at the appropriate time intervals (e.g., once a year), we
may detect time variability in System~C. This observation would have
the additional benefit that we could detect the formation of a BAL in
System~A as we have speculated. Since the equivalent width of System~A
has doubled in only $\sim$ 1.3 years, a few more observations may show
dramatic evolution. We could also diagnose variations in the
ionization state of the absorber by comparing observed spectra with
our simulations. Spectropolarimetric observations would also be
worthwhile because the degree of polarization in the troughs of
intrinsic NALs could be much larger than those in intervening NALs,
indicating scattering around the absorber. We believe that
HS\,1603+3820, one of the best laboratories for the investigation of
intrinsic NALs, is an excellent target to monitor over a long period
of time.

\acknowledgments 
This work was supported by NASA grant NAG5-10817. We are grateful to
the staff of the Subaru telescope, which is operated by the National
Astronomical Observatory of Japan. We would also like to thank
Christopher Churchill for providing us with the {\sc minfit} software
package, and Ben Lackey for his software to evaluate the
pixel-by-pixel covering factors. Finally, we wish to thank the
anonymous referee for many helpful comments and suggestions.

\clearpage

\begin{deluxetable}{rcccccccccccc}
\rotate
\tabletypesize{\scriptsize}
\setlength{\tabcolsep}{0.05in}
\tablecaption{Metal lines detected in 1st and 2nd spectra \label{t1}}
\tablewidth{0pt}
\tablehead{
\colhead{(1)} &
\colhead{(2)} &
\colhead{(3)} &
\colhead{(4)} &
\colhead{(5)} &
\colhead{(6)} &
\colhead{(7)} &
\colhead{(8)} &
\colhead{(9)} &
\colhead{(10)} &
\colhead{(11)} &
\colhead{(12)} &
\colhead{(13)} \\
\colhead{Line ID} & 
\colhead{$\lambda_{obs}$} & 
\colhead{$z_{abs}$} & 
\colhead{V} &
\colhead{b} &
\colhead{$C_{f}$} &
\colhead{$\log N$} &
\colhead{$\lambda_{obs}$} & 
\colhead{$z_{abs}$} & 
\colhead{V} &
\colhead{b} &
\colhead{$C_{f}$} &
\colhead{$\log N$} \\
\colhead{} & 
\colhead{(\AA)} & 
\colhead{} & 
\colhead{(km s$^{-1}$)} & 
\colhead{(km s$^{-1}$)} & 
\colhead{} & 
\colhead{(cm$^{-2}$)} &
\colhead{(\AA)} & 
\colhead{} & 
\colhead{(km s$^{-1}$)} & 
\colhead{(km s$^{-1}$)} & 
\colhead{} & 
\colhead{(cm$^{-2}$)} \\
}
\startdata
\tableline
\multicolumn{13}{c}{System A : $z_{abs}=2.4302$} 
\\
\tableline
 \civ\ $\lambda$1548: &        &        &       &                &                 &             &        &        &       &                &                 &                \\
 1............        & 5292.6 & 2.4185 & 10646 & $86\pm 2^{a}$  &$0.31\pm0.01^{a}$&$14.59\pm0.04^{a}$& 5292.6 & 2.4185 & 10639 &$88\pm 1^{a}$&$0.44\pm0.01^{a}$&$14.62\pm0.01^{a}$ \\
 2............	      &	       &	&	&      	         &                 &      	 & 5298.4 & 2.4222 & 10319 & $406\pm 8^{a}$ & 0.44$^{a}$      &$14.76\pm0.01^{a}$ \\
 3............        & 5301.6 & 2.4246 & 10106 & $238\pm 5^{a}$ &  0.31$^{a}$     &$14.59\pm0.01^{a}$&   &        &       &                &                 &                \\
 4............        & 5316.5 & 2.4340 & 9287  &  60.3$^{b}$    &  ...            & 13.60$^{b}$ & 5316.8 & 2.4342 & 9269  &  70.7$^{b}$    &  ...            & 13.91$^{b}$    \\
 5............        & 5319.8 & 2.4361 & 9104  &  78.5$^{b}$    &  ...            & 13.82$^{b}$ & 5319.9 & 2.4362 & 9095  &  77.7$^{b}$    &  ...            & 14.20$^{b}$    \\
 6............        & 5320.4 & 2.4365 & 9069  &  27.1$^{b}$    &  ...            & 13.45$^{b}$ & 5320.5 & 2.4366 & 9060  &  23.2$^{b}$    &  ...            & 13.36$^{b}$    \\
 7............        & 5322.8 & 2.4381 & 8929  &  50.7$^{b}$    &  ...            & 12.84$^{b}$ & 5322.8 & 2.4381 & 8929  &  48.1$^{b}$    &  ...            & 13.53$^{b}$    \\
 8............	      &	       &	&	&                &                 &             & 5324.1 & 2.4389 & 8859  &  31.5$^{b}$    &  ...            & 13.15$^{b}$    \\
 9............	      & 5326.1 & 2.4402 & 8746  & 289.3$^{b}$    &  ...            & 13.97$^{b}$ & 5325.8 & 2.4400 & 8764  &  84.1$^{b}$    &  ...            & 13.91$^{b}$    \\
 10...........        & 5327.6 & 2.4412 & 8659  &   9.9$^{b}$    &  ...            & 13.23$^{b}$ & 5327.6 & 2.4412 & 8659  &  12.3$^{b}$    &  ...            & 13.54$^{b}$    \\
 11...........        & 5327.6 & 2.4412 & 8659  &  42.1$^{b}$    &  ...            & 13.47$^{b}$ & 5327.6 & 2.4412 & 8659  &  41.1$^{b}$    &  ...            & 13.80$^{b}$    \\
 12...........        & 5329.4 & 2.4423 & 8563  &  38.9$^{b}$    &  ...            & 13.29$^{b}$ & 5329.4 & 2.4423 & 8563  &  51.8$^{b}$    &  ...            & 13.96$^{b}$    \\
 13...........        &        &        &       &                &                 &             & 5330.7 & 2.4432 & 8485  &  24.1$^{b}$    &  ...            & 12.94$^{b}$    \\
 14...........        & 5331.7 & 2.4438 & 8433  &  59.4$^{b}$    &  ...            & 13.34$^{b}$ & 5331.8 & 2.4439 & 8424  &  49.0$^{b}$    &  ...            & 13.70$^{b}$    \\
 15...........        & 5334.2 & 2.4454 & 8293  &  54.8$^{b}$    &  ...            & 13.54$^{b}$ & 5334.3 & 2.4455 & 8285  &  82.1$^{b}$    &  ...            & 13.99$^{b}$    \\
\\			      			      
\tableline		      			      
\multicolumn{13}{c}{System B : $z_{abs}=2.4785$}	      
\\			      			      
\tableline		      			      
 \alii $\lambda$1671: &        &        &       &                &                 &                  &        &        &       &                &                 &                  \\
 1............        & 5811.2 & 2.4781 & 5461  & 11.9 $\pm$ 0.8 & ...             & 11.99 $\pm$ 0.02 & 5811.3 & 2.4782 & 5452  & 10.6 $\pm$ 1.1 & ...             & 11.88 $\pm$ 0.04 \\
 2............        & 5812.0 & 2.4786 & 5418  &  9.3 $\pm$ 0.2 & ...             & 12.88 $\pm$ 0.02 & 5812.0 & 2.4786 & 5418  &  9.3 $\pm$ 0.3 & ...             & 12.81 $\pm$ 0.03 \\
 3............        & 5813.3 & 2.4794 & 5349  &  4.8 $\pm$ 1.6 & ...             & 11.23 $\pm$ 0.08 & 5813.3 & 2.4794 & 5349  &  1.3 $\pm$ 3.5 & ...             & 11.03 $\pm$ 0.14 \\
 4............        & 5813.8 & 2.4797 & 5323  &  5.7 $\pm$ 0.6 & ...             & 11.75 $\pm$ 0.03 & 5813.8 & 2.4797 & 5323  &  4.8 $\pm$ 0.9 & ...             & 11.62 $\pm$ 0.05 \\
 5............        & 5815.2 & 2.4805 & 5254  &  2.9 $\pm$ 0.3 & ...             & 11.95 $\pm$ 0.03 & 5815.2 & 2.4805 & 5254  &  2.8 $\pm$ 0.5 & ...             & 11.91 $\pm$ 0.04 \\
 6............        & 5816.3 & 2.4812 & 5194  &  3.8 $\pm$ 0.7 & ...             & 11.71 $\pm$ 0.05 & 5816.3 & 2.4812 & 5194  &  3.1 $\pm$ 1.2 & ...             & 11.62 $\pm$ 0.10 \\
 7............        & 5816.7 & 2.4814 & 5177  &  6.2 $\pm$ 1.6 & ...             & 11.55 $\pm$ 0.08 & 5816.5 & 2.4813 & 5185  &  5.7 $\pm$ 3.5 & ...             & 11.36 $\pm$ 0.19 \\
 \civ  $\lambda$1548: &        &        &       &                &                 &                  &        &        &       &                &                 &                  \\
 1............        & 5381.8 & 2.4762 & 5627  &  8.5 $\pm$ 0.3$^{c}$ & ...             & 13.36 $\pm$ 0.01$^{c}$ & 5381.8 & 2.4762 & 5625  &  8.6 $\pm$ 0.6       & 1.54 $\pm$ 0.61 & 13.07 $\pm$ 0.21       \\
 2............        & 5382.2 & 2.4764 & 5606  &  6.6 $\pm$ 0.2$^{c}$ & ...             & 14.08 $\pm$ 0.02$^{c}$ & 5382.1 & 2.4764 & 5608  &  7.3 $\pm$ 0.3       & 0.98 $\pm$ 0.02 & 14.06 $\pm$ 0.05       \\
 3............        & 5382.8 & 2.4768 & 5571  & 23.2 $\pm$ 1.0$^{c}$ & ...             & 13.49 $\pm$ 0.02$^{c}$ & 5382.8 & 2.4768 & 5573  & 14.2 $\pm$ 2.0$^{c}$ & ...             & 13.31 $\pm$ 0.06$^{c}$ \\
 4............        & 5383.0 & 2.4770 & 5559  &  2.2 $\pm$ 1.3$^{c}$ & ...             & 12.27 $\pm$ 0.09$^{c}$ & 5383.1 & 2.4770 & 5556  &  3.6 $\pm$ 2.1$^{c}$ & ...             & 12.51 $\pm$ 0.22$^{c}$ \\
 5............        &        &        &       &                      &                 &                        & 5383.4 & 2.4772 & 5539  &  9.7 $\pm$ 3.9$^{c}$ & ...             & 11.51 $\pm$ 9.99$^{c}$ \\
 6............        &        &        &       &                      &                 &                        & 5383.5 & 2.4773 & 5530  & 25.0 $\pm$ 2.0$^{c}$ & ...             & 13.27 $\pm$ 0.23$^{c}$ \\
 7............        & 5383.8 & 2.4775 & 5517  & 16.5 $\pm$ 2.0       & 0.95 $\pm$ 0.03 & 13.67 $\pm$ 0.18       & 5383.8 & 2.4775 & 5513  &  9.0 $\pm$ 0.5       & 0.99 $\pm$ 0.02 & 14.22 $\pm$ 0.05       \\
 8............        & 5383.9 & 2.4775 & 5511  &  7.2 $\pm$ 0.6$^{c}$ & ...             & 14.16 $\pm$ 0.04$^{c}$ &        &        &       &                      &                 &                        \\
 9............        & 5384.5 & 2.4779 & 5477  & 14.9 $\pm$ 5.5       & 1.01 $\pm$ 0.15 & 13.59 $\pm$ 0.09       & 5384.5 & 2.4779 & 5478  & 12.5 $\pm$ 0.9       & 1.00 $\pm$ 0.07 & 13.72 $\pm$ 0.06       \\
 10...........        & 5384.5 & 2.4779 & 5475  &  4.1 $\pm$ 1.4       & 0.85 $\pm$ 0.56 & 13.10 $\pm$ 0.43       &        &        &       &                      &                 &                        \\
 11...........        & 5385.0 & 2.4783 & 5447  & 13.5 $\pm$ 1.4$^{c}$ & ...             & 14.03 $\pm$ 0.05$^{c}$ & 5385.1 & 2.4783 & 5444  & 12.4 $\pm$ 1.4       & 0.98 $\pm$ 0.03 & 14.02 $\pm$ 0.05       \\
 12...........        & 5385.4 & 2.4785 & 5428  &  5.0 $\pm$ 3.0       & 0.99 $\pm$ 0.47 & 13.09 $\pm$ 0.70       & 5385.6 & 2.4786 & 5418  & 12.9 $\pm$ 2.9       & 0.85 $\pm$ 0.08 & 14.00 $\pm$ 0.09       \\
 13...........        & 5385.6 & 2.4787 & 5413  & 12.8 $\pm$ 6.2       & 0.98 $\pm$ 0.01 & 13.99 $\pm$ 0.19       &        &        &       &                      &                 &                        \\
 14...........        & 5386.1 & 2.4790 & 5386  &  7.7 $\pm$ 1.7       & 0.89 $\pm$ 0.07 & 13.93 $\pm$ 0.06       & 5386.2 & 2.4790 & 5383  & 26.1 $\pm$ 2.1       & 0.99 $\pm$ 0.03 & 14.20 $\pm$ 0.06       \\
 15...........        & 5386.3 & 2.4791 & 5378  & 27.2 $\pm$ 0.5$^{c}$ & ...             & 13.90 $\pm$ 0.01$^{c}$ &        &        &       &                      &                 &                        \\
 16...........        & 5387.1 & 2.4796 & 5334  &  6.2 $\pm$ 1.2       & 1.29 $\pm$ 1.50 & 12.73 $\pm$ 0.57       & 5387.3 & 2.4797 & 5323  & 16.4 $\pm$ 1.8       & 0.69 $\pm$ 0.29 & 13.68 $\pm$ 0.26       \\
 17...........        & 5387.4 & 2.4798 & 5317  &  8.2 $\pm$ 0.3       & 0.98 $\pm$ 0.02 & 13.75 $\pm$ 0.03       & 5387.4 & 2.4798 & 5314  &  7.0 $\pm$ 0.9       & 1.02 $\pm$ 0.19 & 13.49 $\pm$ 0.16       \\
 18...........        & 5388.4 & 2.4804 & 5262  & 18.2 $\pm$ 2.4       & 0.59 $\pm$ 0.77 & 13.25 $\pm$ 0.65       & 5388.3 & 2.4804 & 5263  & 11.3 $\pm$ 4.4       & 0.16 $\pm$ 0.02 & 14.05 $\pm$ 0.36       \\
 19...........        & 5388.6 & 2.4806 & 5250  &  4.7 $\pm$ 0.2       & 0.97 $\pm$ 0.02 & 13.97 $\pm$ 0.06       & 5388.6 & 2.4806 & 5246  &  4.7 $\pm$ 0.4       & 0.93 $\pm$ 0.03 & 14.06 $\pm$ 0.15       \\
 \feii $\lambda$1608: &        &        &       &                &                 &                  &        &        &       &                &                 &                  \\
 1............        & 5595.1 & 2.4786 & 5418  &  4.1 $\pm$ 0.5 & ...             & 13.69 $\pm$ 0.03 & 5595.1 & 2.4786 & 5418  &  6.3 $\pm$ 0.5 & ...             & 13.60 $\pm$ 0.03 \\
 \siii $\lambda$1527: &        &        &       &                &                 &                  &        &        &       &                &                 &                  \\
 1............        & 5302.9 & 2.4737 & 5840  & 22.9 $\pm$ 4.8$^{c}$ & ...       & 13.06 $\pm$ 0.07$^{c}$ & 5302.9 & 2.4737 & 5840  & 19.0 $\pm$ 6.2$^{c}$  & ...      & 13.04 $\pm$ 0.06$^{c}$  \\
 2............        & 5303.7 & 2.4742 & 5797  & 21.8 $\pm$ 3.1$^{c}$ & ...       & 13.14 $\pm$ 0.02$^{c}$ & 5303.7 & 2.4742 & 5797  & 23.7 $\pm$ 3.8$^{c}$  & ...      & 13.10 $\pm$ 0.07$^{c}$  \\
 3............        & 5304.6 & 2.4748 & 5746  & 23.6 $\pm$ 4.8$^{c}$ & ...       & 12.95 $\pm$ 0.04$^{c}$ & 5304.6 & 2.4748 & 5746  & 19.1 $\pm$ 4.0$^{c}$  & ...      & 12.95 $\pm$ 0.08$^{c}$  \\
 4............        & 5306.4 & 2.4757 & 5668  & 11.2 $\pm$ 0.1$^{c}$ & ...       & 13.63 $\pm$ 0.01$^{c}$ & 5306.4 & 2.4757 & 5668  & 10.9 $\pm$ 0.8$^{c}$  & ...      & 13.61 $\pm$ 0.03$^{c}$  \\
 5............        & 5307.4 & 2.4765 & 5608  & 13.8 $\pm$ 0.8$^{c}$ & ...       & 13.04 $\pm$ 0.03$^{c}$ & 5307.6 & 2.4765 & 5599  & 12.3 $\pm$ 1.1$^{c}$  & ...      & 13.03 $\pm$ 0.06$^{c}$  \\
 6............        & 5310.2 & 2.4782 & 5452  & 14.2 $\pm$ 1.0$^{c}$ & ...       & 13.24 $\pm$ 0.02$^{c}$ & 5310.0 & 2.4782 & 5461  & 14.2 $\pm$ 2.3$^{c}$  & ...      & 13.24 $\pm$ 0.06$^{c}$  \\
 7............        & 5310.8 & 2.4786 & 5418  & 11.2 $\pm$ 0.1$^{c}$ & ...       & 14.07 $\pm$ 0.02$^{c}$ & 5310.8 & 2.4786 & 5418  & 11.2 $\pm$ 0.3$^{c}$  & ...      & 14.07 $\pm$ 0.03$^{c}$  \\
 8............        & 5312.5 & 2.4798 & 5323  & 12.3 $\pm$ 0.4$^{c}$ & ...       & 13.17 $\pm$ 0.01$^{c}$ & 5312.5 & 2.4798 & 5323  & 12.3 $\pm$ 1.5$^{c}$  & ...      & 13.17 $\pm$ 0.02$^{c}$  \\
 9............        & 5313.7 & 2.4805 & 5254  &  4.7 $\pm$ 0.2$^{c}$ & ...       & 13.45 $\pm$ 0.01$^{c}$ & 5313.9 & 2.4805 & 5246  &  5.6 $\pm$ 0.2$^{c}$  & ...      & 13.41 $\pm$ 0.03$^{c}$  \\
 10...........        & 5314.8 & 2.4812 & 5194  &  7.2 $\pm$ 0.5$^{c}$ & ...       & 13.22 $\pm$ 0.03$^{c}$ & 5314.8 & 2.4812 & 5194  &  7.8 $\pm$ 0.9$^{c}$  & ...      & 13.22 $\pm$ 0.05$^{c}$  \\
 11...........        & 5315.1 & 2.4815 & 5177  & 11.7 $\pm$ 0.4$^{c}$ & ...       & 13.43 $\pm$ 0.02$^{c}$ & 5315.2 & 2.4815 & 5168  & 11.5 $\pm$ 0.4$^{c}$  & ...      & 13.49 $\pm$ 0.02$^{c}$  \\
\\			      			     
\tableline		      			     
\multicolumn{13}{c}{System C : $z_{abs}=2.5370$}      
\\			      			     
\tableline		      			     
 \civ  $\lambda$1548: &        &        &       &                &                 &                  &        &        &       &                &                 &                  \\
 1............        & 5475.8 & 2.5369 & 432   & 11.6 $\pm$ 0.3 & 1.10 $\pm$ 0.17 & 13.30 $\pm$ 0.09 & 5475.8 & 2.5369 & 432   & 11.4 $\pm$ 0.5 & 0.88 $\pm$ 0.16 & 13.44 $\pm$ 0.11 \\ 
\\			      			     
\tableline		      			     
\multicolumn{13}{c}{System D : $z_{abs}=2.5532$}      
\\			      			     
\tableline		      			     
 \civ  $\lambda$1548: &        &        &       &                &                 &                  &        &        &       &                &                 &                  \\
 1............        & 5497.9 & 2.5512 & $-$778&  4.8 $\pm$ 0.7 & 0.37 $\pm$ 0.21 & 13.07 $\pm$ 0.35 &        &        &       &                &                 &                  \\
 2............        & 5500.1 & 2.5526 & $-$896& 11.0 $\pm$ 0.6 & 0.91 $\pm$ 0.35 & 13.15 $\pm$ 0.20 & 5500.1 & 2.5526 & $-$896&  9.5 $\pm$ 1.4 & 0.59 $\pm$ 0.34 & 13.34 $\pm$ 0.35 \\
 3............        & 5500.9 & 2.5531 & $-$939& 13.1 $\pm$ 0.2 & 0.98 $\pm$ 0.01 & 14.28 $\pm$ 0.01 & 5501.0 & 2.5532 & $-$947& 14.4 $\pm$ 1.1 & 0.99 $\pm$ 0.03 & 14.17 $\pm$ 0.07 \\
 4............        & 5501.5 & 2.5535 & $-$972&  8.1 $\pm$ 0.2 & 0.99 $\pm$ 0.00 & 15.05 $\pm$ 0.08 & 5501.5 & 2.5535 & $-$972&  8.9 $\pm$ 1.0 & 0.97 $\pm$ 0.02 & 14.82 $\pm$ 0.25 \\
\enddata
\tablenotetext{a}{Parameters are obtained by manual fitting method
  with \cf\ left as a free parameter. The uncertainties given here are
  the result of Poisson errors only and were evaluated by scanning
  parameter space, as discussed in \S4.1 of the text. The covering
  factors for components 1 and 3 (in the 1st spectrum) and 1 and 2 (in
  the 2nd spectrum) were required to be the same.}
\tablenotetext{b}{Parameters are obtained by manual fitting, assuming
  $C_{f}=1.0$. The uncertainties are discussed in \S4.1 of the text.}
\tablenotetext{c}{Parameters are obtained by automatic fitting,
  assuming $C_{f}=1.0$. The errors quoted here reflect only the
  uncertainties due to Poisson noise.}
\end{deluxetable}

\begin{deluxetable}{rcccccc}
\tabletypesize{\scriptsize}
\setcounter{table}{1}
\tablecaption{Metal lines in the 2nd spectrum \label{t2}}
\tablewidth{0pt}
\tablehead{
\colhead{(1)} &
\colhead{(2)} &
\colhead{(3)} &
\colhead{(4)} &
\colhead{(5)} &
\colhead{(6)} &
\colhead{(7)} \\
\colhead{} & 
\colhead{} & 
\colhead{} & 
\colhead{} & 
\colhead{} & 
\colhead{} & 
\colhead{} \\
\colhead{Line ID} & 
\colhead{$\lambda_{obs}$} & 
\colhead{$z_{abs}$} & 
\colhead{V} &
\colhead{b} &
\colhead{$C_{f}$} &
\colhead{$\log N$} \\
\colhead{} & 
\colhead{(\AA)} & 
\colhead{} & 
\colhead{(km s$^{-1}$)} & 
\colhead{(km s$^{-1}$)} & 
\colhead{} & 
\colhead{(cm$^{-2}$)} \\
}
\startdata
\tableline		      			      
\multicolumn{7}{c}{System B : $z_{abs}=2.4785$}       
\\			      			      
\tableline		      			      
 \cii  $\lambda$1335: &        &        &       &                &                 &                  \\
 1............        & 4639.4 & 2.4764 & 5608  &  4.8 $\pm$ 0.9 & ...             & 13.28 $\pm$ 0.06 \\
 2............        & 4641.6 & 2.4781 & 5461  & 13.0 $\pm$ 1.4 & ...             & 14.04 $\pm$ 0.05 \\
 3............        & 4642.2 & 2.4785 & 5427  & 10.4 $\pm$ 3.3 & ...             & 14.28 $\pm$ 0.13 \\
 4............        & 4642.4 & 2.4787 & 5409  &  5.4 $\pm$ 1.8 & ...             & 13.87 $\pm$ 0.23 \\
 5............        & 4642.8 & 2.4790 & 5383  & 17.8 $\pm$ 9.7 & ...             & 13.09 $\pm$ 0.17 \\
 6............        & 4643.5 & 2.4795 & 5340  &  6.7 $\pm$ 2.1 & ...             & 12.97 $\pm$ 0.10 \\
 7............        & 4643.8 & 2.4797 & 5323  &  6.0 $\pm$ 0.6 & ...             & 13.75 $\pm$ 0.06 \\
 8............        & 4644.8 & 2.4805 & 5254  &  6.0 $\pm$ 0.6 & ...             & 13.96 $\pm$ 0.10 \\
 9............        & 4645.8 & 2.4812 & 5194  &  4.0 $\pm$ 1.8 & ...             & 14.25 $\pm$ 0.66 \\
 10...........        & 4645.9 & 2.4813 & 5185  &  4.4 $\pm$ 1.5 & ...             & 13.76 $\pm$ 0.15 \\
 11...........        & 4646.3 & 2.4816 & 5159  &  5.6 $\pm$ 2.3 & ...             & 12.82 $\pm$ 0.12 \\
 \nv  $\lambda$1239:  &        &        &       &                &                 &                  \\
 1............        & 4308.0 & 2.4775 & 5513  &  7.9 $\pm$ 1.3       & 0.70 $\pm$ 0.49 & 13.41 $\pm$ 0.49       \\
 2............        & 4308.5 & 2.4779 & 5478  &  7.7 $\pm$ 4.3$^{a}$ & ...             & 12.48 $\pm$ 0.21$^{a}$ \\
 3............        & 4308.9 & 2.4782 & 5452  & 13.5 $\pm$ 4.0$^{a}$ & ...             & 12.90 $\pm$ 0.11$^{a}$ \\
 4............        & 4309.2 & 2.4785 & 5427  &  7.7 $\pm$ 1.1$^{a}$ & ...             & 12.69 $\pm$ 0.05$^{a}$ \\
 5............        & 4309.5 & 2.4787 & 5409  &            ...$^{b}$ &       ...$^{b}$ &              ...$^{b}$ \\
 6............        & 4309.9 & 2.4790 & 5383  &  9.0 $\pm$ 1.5$^{a}$ & ...             & 12.65 $\pm$ 0.12$^{a}$ \\
 \oi  $\lambda$1302:  &        &        &       &                &                 &                  \\
 1............        & 4529.6 & 2.4785 & 5427  &  8.3 $\pm$ 0.3 & ...             & 14.34 $\pm$ 0.03 \\
 2............        & 4533.1 & 2.4812 & 5194  & 12.2 $\pm$ 5.4 & ...             & 13.43 $\pm$ 0.15 \\
 \siii  $\lambda$1260:&        &        &       &                &                 &                  \\
 1............        & 4381.4 & 2.4761 & 5634  &  4.6 $\pm$ 1.5 & ...             & 12.03 $\pm$ 0.08 \\
 2............        & 4381.7 & 2.4764 & 5608  &  3.1 $\pm$ 0.7 & ...             & 12.49 $\pm$ 0.08 \\
 3............        & 4383.6 & 2.4779 & 5478  &  3.2 $\pm$ 1.4 & ...             & 12.18 $\pm$ 0.11 \\
 4............        & 4383.9 & 2.4781 & 5461  & 10.7 $\pm$ 1.2 & ...             & 13.04 $\pm$ 0.03 \\
 5............        & 4384.4 & 2.4785 & 5427  &  2.6 $\pm$ 8.1 & ...             & 15.96 $\pm$ 0.49 \\
 6............        & 4384.6 & 2.4787 & 5409  & 10.2 $\pm$ 3.6 & ...             & 12.90 $\pm$ 0.31 \\
 7............        & 4385.5 & 2.4794 & 5349  &  6.4 $\pm$ 1.6 & ...             & 12.11 $\pm$ 0.08 \\
 8............        & 4385.9 & 2.4797 & 5323  &  5.3 $\pm$ 0.5 & ...             & 12.82 $\pm$ 0.05 \\
 9............        & 4386.9 & 2.4805 & 5254  &  5.4 $\pm$ 0.4 & ...             & 13.02 $\pm$ 0.06 \\
 10...........        & 4387.8 & 2.4812 & 5194  &  4.9 $\pm$ 0.7 & ...             & 13.09 $\pm$ 0.10 \\
 11...........        & 4388.0 & 2.4814 & 5177  &  3.2 $\pm$ 0.9 & ...             & 12.41 $\pm$ 0.07 \\
 \siiii $\lambda$1207:&        &        &       &                &                 &                  \\
 1............        & 4194.9 & 2.4769 & 5564  & 28.4 $\pm$ 4.5 & ...             & 14.34 $\pm$ 0.38 \\
 2............        & 4196.1 & 2.4779 & 5478  & 55.3 $\pm$35.3 & ...             & 13.79 $\pm$ 0.24 \\
 3............        & 4196.9 & 2.4786 & 5418  &  8.8 $\pm$ 8.9 & ...             & 14.05 $\pm$ 2.42 \\
 4............        & 4197.4 & 2.4790 & 5383  & 12.4 $\pm$ 5.2 & ...             & 12.75 $\pm$ 0.30 \\
 5............        & 4198.0 & 2.4795 & 5340  &  7.3 $\pm$ 3.4 & ...             & 12.95 $\pm$ 0.28 \\
 6............        & 4198.3 & 2.4797 & 5323  & 32.7 $\pm$ 6.0 & ...             & 13.10 $\pm$ 0.14 \\
 7............        & 4198.3 & 2.4797 & 5323  &  2.6 $\pm$ 0.4 & ...             & 14.50 $\pm$ 0.68 \\
 8............        & 4199.2 & 2.4805 & 5254  &  7.4 $\pm$ 1.2 & ...             & 13.24 $\pm$ 0.24 \\
 9............        & 4200.1 & 2.4812 & 5194  &  4.1 $\pm$ 1.4 & ...             & 12.48 $\pm$ 0.11 \\
 10...........        & 4200.3 & 2.4814 & 5177  &  5.2 $\pm$ 1.9 & ...             & 12.34 $\pm$ 0.11 \\
 \siiv $\lambda$1394: &        &        &       &                &                 &                  \\
 1............        & 4845.1 & 2.4763 & 5616  & 12.7 $\pm$ 5.7 & ...             & 12.10 $\pm$ 0.34 \\
 2............        & 4845.2 & 2.4764 & 5608  &  4.5 $\pm$ 0.5 & ...             & 13.10 $\pm$ 0.05 \\
 3............        & 4846.8 & 2.4775 & 5513  &  6.6 $\pm$ 0.8 & ...             & 12.43 $\pm$ 0.04 \\
 4............        & 4847.3 & 2.4779 & 5478  &  3.3 $\pm$ 0.6 & ...             & 12.66 $\pm$ 0.05 \\
 5............        & 4847.8 & 2.4782 & 5452  & 20.6 $\pm$ 3.2 & ...             & 13.01 $\pm$ 0.05 \\
 6............        & 4848.3 & 2.4786 & 5418  &  9.6 $\pm$ 0.8 & ...             & 13.08 $\pm$ 0.04 \\
 7............        & 4848.9 & 2.4790 & 5383  & 14.4 $\pm$ 1.4 & ...             & 12.89 $\pm$ 0.03 \\
 8............        & 4849.7 & 2.4796 & 5332  & 24.8 $\pm$ 2.3 & ...             & 12.99 $\pm$ 0.04 \\
 9............        & 4850.0 & 2.4798 & 5314  &  4.6 $\pm$ 0.9 & ...             & 12.74 $\pm$ 0.07 \\
 10...........        & 4851.0 & 2.4805 & 5254  &  5.8 $\pm$ 0.5 & ...             & 13.17 $\pm$ 0.06 \\
\\			      			      
\tableline		      			      
\multicolumn{7}{c}{System D : $z_{abs}=2.5532$}       
\\			      			      
\tableline		      			      
 \siiv $\lambda$1394: &        &        &       &                &                 &                  \\
 1............        & 4952.2 & 2.5531 & $-$939& 10.0 $\pm$ 1.1 & 0.77 $\pm$ 0.42 & 12.91 $\pm$ 0.30 \\
 2............        & 4952.7 & 2.5535 & $-$972&  8.5 $\pm$ 0.4 & 0.94 $\pm$ 0.07 & 13.39 $\pm$ 0.07 \\
\\			      			      
\tableline		      			      
\multicolumn{7}{c}{System E : $z_{abs}=1.8875$}       
\\			      			      
\tableline		      			      
 \civ  $\lambda$1548: &        &        &       &                &                 &                  \\
 1............        & 4470.1 & 1.8873 & 60471 & 27.1 $\pm$ 2.9 & 0.95 $\pm$ 0.28 & 13.80 $\pm$ 0.17 \\
 2............        & 4470.9 & 1.8878 & 60422 & 16.1 $\pm$ 3.2 & 0.50 $\pm$ 0.23 & 13.70 $\pm$ 0.30 \\
\\			      			      
\tableline		      			      
\multicolumn{7}{c}{System F : $z_{abs}=1.9644$}       
\\			      			      
\tableline		      			      
 \civ  $\lambda$1548: &        &        &       &                &                 &                  \\
 1............        & 4587.3 & 1.9630 & 52985 & 11.2 $\pm$ 2.3 & 0.47 $\pm$ 0.19 & 13.50 $\pm$ 0.26 \\
 2............        & 4587.8 & 1.9633 & 52956 &  7.1 $\pm$ 0.7 & 0.96 $\pm$ 0.05 & 13.86 $\pm$ 0.07 \\
 3............        & 4587.9 & 1.9634 & 52946 &  7.7 $\pm$ 0.5 & 0.98 $\pm$ 0.06 & 13.76 $\pm$ 0.06 \\
 4............        & 4588.5 & 1.9638 & 52907 &  3.0 $\pm$ 2.6 & 0.11 $\pm$ 0.02 & 15.82 $\pm$ 3.57 \\
 5............        & 4589.5 & 1.9644 & 52848 &  8.3 $\pm$ 0.5 & 0.96 $\pm$ 0.03 & 13.98 $\pm$ 0.06 \\
 6............        & 4589.8 & 1.9646 & 52828 & 11.7 $\pm$ 3.3 & 0.69 $\pm$ 0.40 & 13.34 $\pm$ 0.33 \\
 7............        & 4590.2 & 1.9649 & 52799 &  8.6 $\pm$ 1.1 & 0.76 $\pm$ 0.15 & 13.48 $\pm$ 0.13 \\
 8............        & 4590.4 & 1.9650 & 52789 &  7.2 $\pm$ 1.5 & 0.73 $\pm$ 0.69 & 13.06 $\pm$ 0.51 \\
\\			      			      
\tableline		      			      
\multicolumn{7}{c}{System G : $z_{abs}=2.0701$}       
\\			      			      
\tableline		      			      
 \civ  $\lambda$1548: &        &        &       &                &                 &                  \\
 1............        & 4752.5 & 2.0697 & 42643 &  6.6 $\pm$ 1.1 & 1.29 $\pm$ 0.40 & 13.38 $\pm$ 0.23 \\
 2............        & 4753.0 & 2.0700 & 42614 & 19.4 $\pm$ 1.8 & 1.04 $\pm$ 0.05 & 14.16 $\pm$ 0.05 \\
 \siiv $\lambda$1403: &        &        &       &                &                 &                  \\
 1............        & 4306.2 & 2.0698 & 42633 & 13.2 $\pm$ 2.5 & ...             & 13.10 $\pm$ 0.06 \\
 2............        & 4306.6 & 2.0701 & 42604 &  6.8 $\pm$ 2.1 & ...             & 12.75 $\pm$ 0.11 \\
\\
\tableline		      			      
\multicolumn{7}{c}{System H : $z_{abs}=2.2653$}       
\\			      			      
\tableline
 \civ  $\lambda$1548: &        &        &       &                &                 &                  \\
 1............        & 5055.2 & 2.2652 & 24357 &  5.8 $\pm$ 1.8 & 0.49 $\pm$ 0.30 & 13.23 $\pm$ 0.33 \\
 2............        & 5055.5 & 2.2654 & 24339 &      ...$^{c}$ &       ...$^{c}$ &        ...$^{c}$ \\
 \siiii $\lambda$1207:&        &        &       &                &                 &                  \\
 1............        & 3939.5 & 2.2652 & 24357 &  3.6 $\pm$ 5.6 & ...             & 13.02 $\pm$ 2.41 \\
\enddata
\tablenotetext{a}{Parameters are obtained by automatic fitting,
  assuming $C_{f}=1.0$.}
\tablenotetext{b}{Component 5 could not be fit for a continuum that
  gives physical solution for the other components. \nv\ $\lambda$1243
  could be contaminated by a blend or affected by correlated noise in
  this region.}
\tablenotetext{c}{It appears that component 2 of \civ\ $\lambda$1548
  is affected by a poor continuum fit of by contamination since \civ\
  $\lambda$1551 is not detected at the same velocity, so no physical
  solution can be obtained.}
\end{deluxetable}

\begin{deluxetable}{cccccc}
\tablecaption{Error Bar on $C_f$ from Continuum Fitting Uncertainty$^{a}$ \label{t3}}
\tablewidth{0pt}
\tablehead{
\colhead{(1)} &
\colhead{(2)} &
\colhead{(3)} &
\colhead{(4)} &
\colhead{(5)} &
\colhead{(6)} \\
\noalign{\vskip 6pt}
\colhead{} &
\multicolumn{5}{c}{$R_b$\tablenotemark{b}} \\
\cline{2-6} \\
\noalign{\vskip -10pt}
\colhead{$C_f$} &
\colhead{0.1} &
\colhead{0.3} &
\colhead{0.5} &
\colhead{0.7} &
\colhead{0.9} 
}
\startdata
0.0 & ...             & ...                & ...                & ...                & ...               \\ \\
0.1 & ...             & ...                & ...                & ...                & $<\pm 0.01$       \\ \\
0.2 & ...             & ...                & ...                & ...                & $^{+0.8}_{-0.09}$ \\ \\
0.3 & ...             & ...                & ...                & $<\pm 0.01$        & $^{+0.7}_{-0.18}$ \\ \\
0.4 & ...             & ...                & ...                & $^{+0.09}_{-0.04}$ & $^{+0.6}_{-0.28}$ \\ \\
0.5 & ...             & ...                & $<\pm 0.01$        & $^{+0.26}_{-0.10}$ & $^{+0.5}_{-0.37}$ \\ \\
0.6 & ...             & ...                & $^{+0.04}_{-0.03}$ & $^{+0.4}_{-0.15}$  & $^{+0.4}_{-0.47}$ \\ \\
0.7 & ...             & $<\pm 0.01$        & $^{+0.09}_{-0.06}$ & $^{+0.3}_{-0.22}$  & $^{+0.3}_{-0.57}$ \\ \\
0.8 & ...             & $^{+0.03}_{-0.02}$ & $^{+0.17}_{-0.10}$ & $^{+0.2}_{-0.29}$  & $^{+0.2}_{-0.67}$ \\ \\
0.9 & $<\pm 0.01$     & $^{+0.06}_{-0.04}$ & $^{+0.1}_{-0.14}$  & $^{+0.1}_{-0.36}$  & $^{+0.1}_{-0.78}$ \\ \\
1.0 & $^{+0}_{-0.02}$ & $^{+0}_{-0.07}$    & $^{+0}_{-0.18}$    & $^{+0}_{-0.44}$    & $^{+0}_{-0.87}$   \\
\enddata
\tablenotetext{a}{Blank entries indicate unphysical combinations of \cf\ and 
                  $R_b$.}
\tablenotetext{b}{$R_b$ is the normalized residual intensity of the blue
                  member of the doublet, relative to the {\it effective} 
                  continuum}
\end{deluxetable}

\end{document}